\documentclass[twocolumn,longbibliography,prx,aps,superscriptaddress,flushbottom,showpacs]{revtex4-1}
\usepackage{xspace,amsmath,amsfonts,amsthm,amssymb,amsbsy,graphicx,color,enumerate}

\begin{document}

\newcommand{\ms}[1]{\mbox{\scriptsize #1}}
\newcommand{\msb}[1]{\mbox{\scriptsize $\mathbf{#1}$}}
\newcommand{\msi}[1]{\mbox{\scriptsize\textit{#1}}}
\newcommand{\nn}{\nonumber} 
\newcommand{\dg}{^\dagger}
\newcommand{\smallfrac}[2]{\mbox{$\frac{#1}{#2}$}}
\newcommand{\ket}[1]{| {#1} \ra}
\newcommand{\bra}[1]{\la {#1} |}
\newcommand{\pfpx}[2]{\frac{\partial #1}{\partial #2}}
\newcommand{\dfdx}[2]{\frac{d #1}{d #2}}
\newcommand{\half}{\smallfrac{1}{2}}
\newcommand{\s}{{\mathcal S}}
\newcommand{\red}{\color{red}}
\newcommand{\blue}{\color{blue}}
\newtheorem{theo}{Theorem} \newtheorem{lemma}{Lemma}

\title{Accurate Lindblad-form master equation for weakly-damped quantum systems \\ across all regimes}

\author{Gavin McCauley} 
\affiliation{U.S. Army Research Laboratory, Computational and Information Sciences Directorate, Adelphi, Maryland 20783, USA} 
\author{Benjamin Cruikshank} 
\affiliation{U.S. Army Research Laboratory, Computational and Information Sciences Directorate, Adelphi, Maryland 20783, USA} 
\affiliation{Department of Physics, University of Massachusetts at Boston, Boston, MA 02125, USA} 
\author{Denys I. Bondar} 
\affiliation{Department of Physics and Engineering Physics, Tulane University, New Orleans, LA 70118, USA} 
\author{Kurt Jacobs} 
\affiliation{U.S. Army Research Laboratory, Computational and Information Sciences Directorate, Adelphi, Maryland 20783, USA} 
\affiliation{Department of Physics, University of Massachusetts at Boston, Boston, MA 02125, USA} 
\affiliation{Hearne Institute for Theoretical Physics, Louisiana State University, Baton Rouge, LA 70803, USA} 

\begin{abstract} 
Realistic models of quantum systems must include dissipative interactions with an environment. For weakly-damped systems the Lindblad-form Markovian master equation is invaluable for this task due to its tractability and efficiency. This equation only applies, however, when the frequencies of any subset of the system's transitions are either equal (degenerate), or their differences are much greater than the transitions' linewidths (far-detuned). Outside of these two regimes the only available efficient description has been the Bloch-Redfield (B-R) master equation, the efficacy of which has long been controversial due to its failure to guarantee the positivity of the density matrix. The ability to efficiently simulate weakly-damped systems across all regimes is becoming increasingly important, especially in the area of quantum technologies. 
Here we solve this long-standing problem. We discover that a condition on the slope of the spectral density is sufficient to derive a Lindblad form master equation that is accurate for all regimes. We further show that this condition is necessary for weakly-damped systems to be described by the B-R equation or indeed any Markovian master equation. We thus obtain a replacement for the B-R equation over its entire domain of applicability  that is no less accurate, simpler in structure, completely positive, allows simulation by efficient quantum trajectory methods, and unifies the previous Lindblad master equations. We also show via exact simulations that the new master equation can describe systems in which slowly-varying transition frequencies cross each other during the evolution. System identification tools, developed in systems engineering, play an important role in our analysis. We expect these tools to prove useful in other areas of physics involving complex systems. 
\end{abstract} 

\maketitle

\section{Introduction} 
\label{intro}

Weakly damped open systems are important across a wide range of areas in both physics and chemistry, from quantum thermodynamics~\cite{Vin16, Horowitz15, Shizume95} to the control of chemical reactions~\cite{Brif10}, to quantum technologies~\cite{MacFarlane03, Taylor16, Degen17, Dunjko17, Braun18, Crosson16, Farhi11, Reiserer15}. So long as the thermal environment that induces the weak damping has a high cut-off frequency (something we assume throughout), Lindblad-form Markovian master equations are tremendously useful for modelling these systems as they avoid the computationally expensive, and often prohibitive, task of simulating the thermal environment~\cite{Prior10, Chin10}. However, recent developments have made it clear that the regime that existing Lindblad master equations cannot describe --- the ``near degenerate'' regime in which non-degenerate transition frequencies are close together --- while long ignored, is crucial for investigating important questions in a range of topics, including reservoir engineering and cascaded systems~\cite{Stannigel12, Tomadin12, Chang12, Cook18, Goerz18, Vuglar2018}, adiabatic computation~\cite{Albash12, Amin08}, super and sub-radiance~\cite{Dantan06, Clark03, Gross82, Kien05}, and ``weak lasing''~\cite{Aleiner12, Eastham16}, with the possibility that this regime will also reveal new tools for controlling quantum systems.  

Weakly-damped quantum systems can be divided into three regimes depending on the frequency difference between pairs of transitions. These regimes are \textit{degenerate} (the frequency difference is zero), \textit{non-degenerate} (the frequency difference is much greater than the transitions' linewidths), and \textit{near-degenerate} (everything else). The degenerate and non-degenerate regimes are described, respectively, by two quite different Lindblad master equations~\cite{Breuer07, Jacobs14} (for ease of reference we present these master equations in the supplement). The difference between them is exemplified by the fact that degenerate transitions exhibit super and sub-radiance, whereas non-degenerate transitions do not. These two Lindblad master equations are obtained from the Bloch-Redfield master equation by making the secular (rotating-wave) approximation. However, no Lindblad master equation has been obtained for the near-degenerate regime~\cite{Majenz13, Santra17}. Thus to simulate systems in which two or more distinct transitions are separated by less than a few linewidths, one must resort to the Bloch-Redfield (B-R) master equation~\cite{Redfield57, Bloch57}. This equation has long been the subject of debate because it is not guaranteed to preserve the positivity of the density matrix~\cite{Dumcke1979, Gardiner10}. In some subfields (e.g. photo-chemistry~\cite{Zhang1998, Yang2002}), the B-R master equation is used as the standard vehicle for treating weakly-damped systems. Practitioners in other fields, for example quantum optics and many areas of quantum technologies, do not use it because its failure to ensure such a fundamental property as positivity is seen as an indication that it cannot be trusted. 

There have been a number of papers, some quite recent, arguing that the B-R equation is a valid and effective model so long as the system is close to Markovian~\cite{Whitney08, Jeske15, Eastham16}. In particular, Eastham \textit{et al.\ }\cite{Eastham16} considered a model of two coupled linear oscillators that can be solved exactly, and examined how well the B-R equation describes the near-degenerate regime (since this is the regime in which it is needed). They found that the B-R equation was both very accurate and preserved positivity to a very good approximation. They attributed this to the fact that the dynamics of the coupled oscillators stays close to Markovian, which was in turn due to the relatively slow variation of the spectral density. In~\cite{Jeske15} Jeske \textit{et al.\ }also noted that when transitions are close enough that they share the same value of the spectral density, the B-R equation reduces to the degenerate master equation, which is a key element in our analysis here. These recent works raise an interesting question: other authors have assumed that the near-degenerate regime is non-Markovian, due to the apparent lack of a Lindblad-form master equation in that regime~\cite{Majenz13, Santra17}. If the dynamics of weakly-damped systems is indeed Markovian in the near-degenerate regime, then it is not unreasonable to suggest that there may be a completely positive Markovian master equation that accurately describes it.

Here we show that there is a single, Lindblad-form master equation that describes weakly-damped systems across all three regimes. This master equation applies to the Ohmic spectrum, and any spectrum that varies sufficiently slowly on the scale of the Lamb shifts and linewidths. For baths with spectra that satisfy this ``slow variation" condition, our master equation agrees to very high accuracy with the B-R equation, something that follows from our derivation and is confirmed by numerical simulations. We further show, using exact simulations, that when our ``slow variation" condition is broken not only does the B-R equation break down, but so do all Markovian master equations. We thus show that the B-R equation cannot be trusted outside the regime in which our master equation is valid, and in this sense our master equation is a complete replacement for the B-R equation. The Linblad master equation has a simpler form that the Bloch-Redfield equation, and thus provides new insight into the behavior of the near-degenerate regime.

Numerical simulations reveal that our master equation describes non-degenerate transitions more accurately than the existing Lindblad master equation for non-degenerate transitions. Not only does our master equation finally provide a non-controversial method for simulating all weakly damped systems; being in the Lindblad form it can also be simulated using efficient Monte Carlo methods~\cite{Wiseman93,Molmer93,Diosi86,Jacobs10b}, and provides a formulation of the action of a thermal bath as a continuous measurement on the system. This quantifies the way in which information flows from the system to the bath. 

We expect that many important problems involving the near-degenerate regime will also involve transition frequencies that change with time, and possibly cross during the evolution. Examples of this are the Landau-Zener transition~\cite{Wubs06, Wittig05} and the control of super- and sub-radiance by shifting energy levels. We show that the adiabatic extension of our master equation is able to accurately describe such time-dependent problems, so long as the rate of change of the transition frequencies is not too fast.   

Our derivation of the Lindblad master equation provides the following corollaries: 
\begin{enumerate}
 
        \item It shows that the secular approximation is unnecessary: weak damping, a high bath cut-off frequency, and sufficient flatness of the spectral density suffice to guarantee positivity and Markovianity. We show that flatness of the spectral density is also a necessary condition for Markovianity 
        
        \item It resolves the controversy regarding the Bloch-Redfield master equation: it shows that when the spectral density is sufficiently flat, this equation is very close to a Lindblad master equation, and will thus approximately preserve positivity. Conversely, outside this flatness condition the Bloch-Redfield equation is, in general, no longer valid, confirming the conjecture in~\cite{Eastham16}).   
\end{enumerate}


We obtain the new master equation in two steps. First, we use exact simulations of a V system coupled to an Ohmic bath, together with the method of system identification, developed in systems engineering, to show that weakly-damped quantum systems with an Ohmic spectrum are not only Markovian but also time-independent across all three regimes. This method also allows us to directly back-out the Lindblad-form equation of motion for this V system. Second, aided by the form obtained in step one, we show how to derive the new Lindblad master equation from the Bloch-Redfield equation valid for all regimes and all temperatures. We provide additional confirmation of its accuracy for the Ohmic bath by comparing its predictions to those of exact simulations for two further systems, a trident system and two co-located qubits. 


In the next section we perform simulations of a V system with an Ohmic bath, and obtain an accurate master equation for this system for all regimes using system identification. In Section~\ref{derivME} we use the information obtained in Section~\ref{SID} to derive a master equation for all weakly damped systems and all temperatures given a constraint on the derivative of the spectral density. In Section~\ref{accnum} we further confirm the accuracy of the master equation with numerical simulations. In Section~\ref{t_apps} we use numerical simulations to show that the master equation is able to describe time-dependent systems whose levels cross. Section~\ref{conc} concludes with a discussion of some open questions.  

\begin{figure}[t] 
\centering
\leavevmode\includegraphics[width=1\hsize]{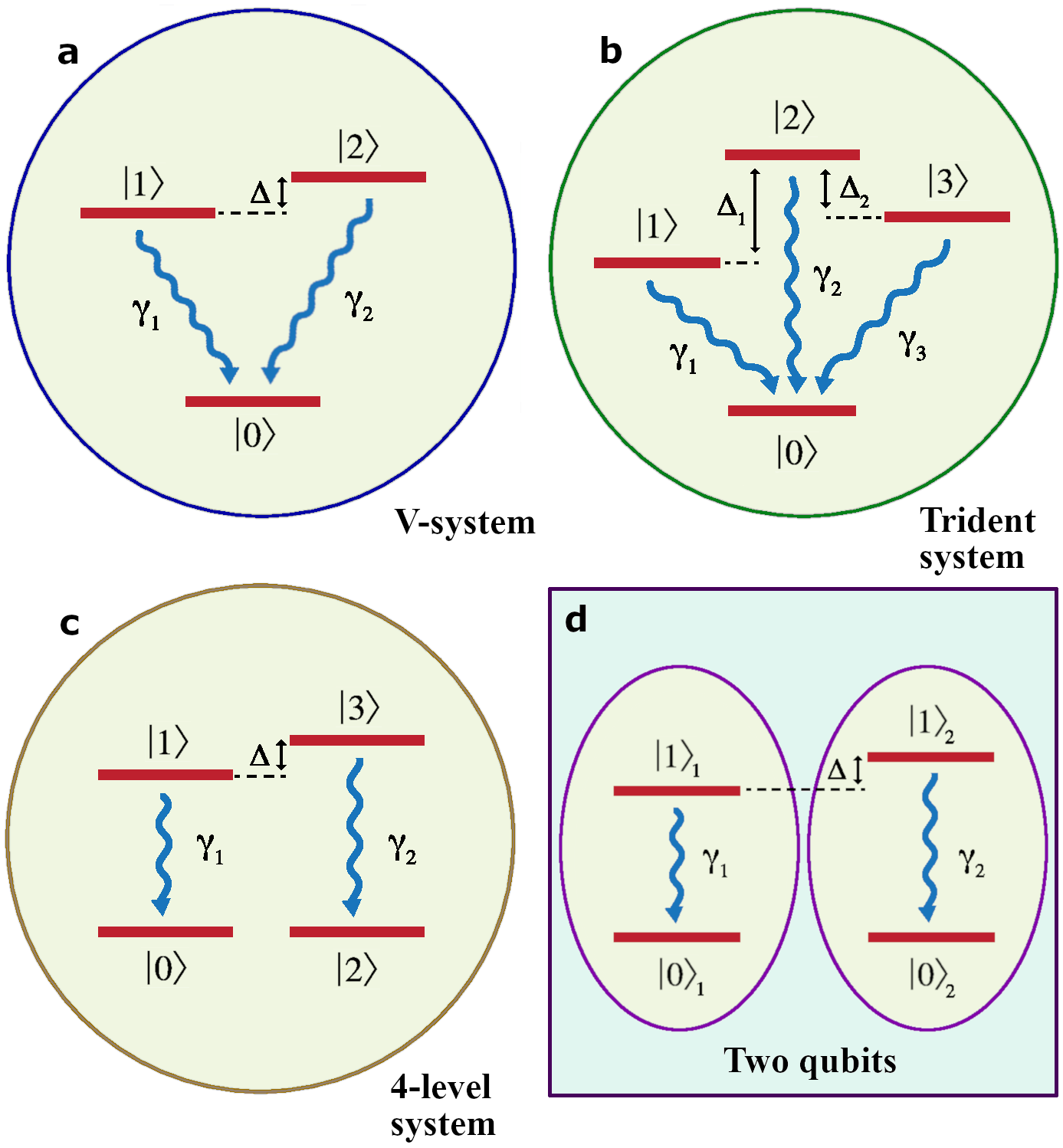}
\caption{(Color online) Here we depict four systems with transitions that decay due to a coupling with a thermal bath at zero temperature. The red bars are the energy eigenstates of the system and the blue wiggly lines indicate the transitions. These system are (a) V system, (b) trident system, (c) four-level system, and (d) to co-located qubits. Given that the relative energy of each level in the diagram is indicated by its vertical position, $\Delta$ denotes the detuning between the transitions in (a), (c), and (d). In system (b) there are three transitions and thus two independent detunings denoted by $\Delta_1$ and $\Delta_2$. The decay rate of the $j^{\ms{th}}$ transition is denoted by $\gamma_j$. The transition operators and frequencies for each of these systems are given in the supplement~\cite{Note3}.}
\label{6systems}
\end{figure}

\section{System identification in the near-degenerate regime}
\label{SID}

The methods of system identification provide us with a way to determine, from the time series of a linear time-invariant system, the minimal number of variables required to generate this time-series (that is, the dimension of the system), as well as its equations of motion. System identification (SID) methods are typically concerned with input/output systems. SID involves obtaining the outputs of a system for a large enough set of distinct inputs that the equations of motion can be determined. While our system does not have inputs, SID methods are easily adapted to replace the set of inputs with a set of initial states. (We give the details of the SID method that we use in Appendix~\ref{App:SID}.) Since evolution of the V system is non-trivial only when the upper levels are populated, and the evolution does not generate coherence with the lower level, the two upper populations together with their complex coherence form a closed four-dimensional system. Performing exact simulations of the V system, SID provides us with the dynamics of a fictitious (and possibly larger) system that generates the four-dimensional dynamics. Specifically, if we denote the state of the fictitious system at time $t$ by $\mathbf{v}(t)$, then SID provides us with a matrix $M(\tau)$ where $\mathbf{v}(\tau) = M(\tau) \mathbf{v}(0)$ for a specified time $\tau$. The number of appreciable eigenvalues of $M$ is the effective size of the fictitious system. 

Since it is only the ratios between the rate parameters that determine the dynamical behavior (up to a scaling of time) we specify all frequencies in terms of an arbitrary frequency, $\tilde{\nu}$. We perform SID on the V system depicted in Fig.\ref{6systems}a with bath cut-off frequency $\Omega = 80\pi$ (details of the bath model are given below in  Section~\ref{derivME}), fix the mean transition frequency $\bar{\omega} \equiv (\omega_1 + \omega_2)/2 = 3 \pi \tilde{\nu}$, and choose the coupling constants $g_1$ and $g_2$ (defined in Eq.(\ref{Adef})) so as to give the decay rates $\gamma_1 = 0.1\tilde{\nu}$ and $\gamma_2 = 0.05\tilde{\nu}$~\footnote{Since we are simulating the Hamiltonian in Eq.(\ref{sysbath}), it is the coupling constants $g_j$, rather than the damping rates $\gamma_j$, that define the simulation, with $\gamma_j \propto |g_j|^2$. Because we have fixed $\bar{\omega}$ while changing the detuning, and since $\gamma_j \propto \omega_j$ for the Ohmic spectral density, when the detuning changes we must change $g_j$ to keep the damping rates fixed}. Since we wish to examine the evolution when the detuning, $\Delta\omega \equiv \omega_2 - \omega_1$, is not large compared to the damping rates, we simulate the evolution for the following four values of $\Delta\omega$: $0$, $0.28\pi \gamma_1$, $2\pi \gamma_1$, and $4.8 \pi \gamma_1$. 

To perform the exact simulations we use the method detailed in~\cite{Prior10, Chin10} which employs the matrix-product-state method of Vidal~\cite{Vidal04, Vidal03}. This in turn requires a split operator method, for which we use a second-order method valid for time-dependent systems, and choose a time-step small enough to obtain an accuracy of about six digits of precision. 

Obtaining the matrix $M$ for each value of the detuning, $\Delta\omega$, we find that the largest four eigenvalues of $M$ account for almost all of the dynamical behavior for all four values: the magnitudes of all the rest of the eigenvalues contribute a fraction of less than $3\times 10^{-4}$ to the 1-norm of $M$. This result implies that the dynamics of the system in the near-degenerate regime is both time-independent and Markovian to very good approximation. 

A 4-dimensional dynamical model for the four independent variables of the V system can now be obtained merely by taking the log of the matrix $M(t)$ for some appropriate value of $t$~\footnote{The time index $t$ must be smaller than the smallest period of the dynamics to avoid multiple branches of the complex logarithm.}. Writing the four variables as the vector $\mathbf{x}$, the approximate model is $\dot{\mathbf{x}} = D \mathbf{x}$, with $D = \ln [M(t)]/t$. To determine the Lindblad-form master equation specified by this model, we need to translate from the elements of $D$ to the familiar terms used to express such master equations. The simplest way to do this is to take a general degenerate master equation for a V system and derive its $D$ matrix. The degenerate master equation for a V system (Fig.~\ref{6systems}a), in which both transitions have frequency $\omega$, is given by 
\begin{align}
   \dot\rho & = -\frac{i}{\hbar}\left[H_0 + H_{\ms{L}},\rho \right] - \mathcal{D}[\Sigma]\rho . 
   \label{degME}
\end{align}
Here  
\begin{align}
 H_0 & = \hbar \omega_0 (|1\rangle \langle 1| + |2\rangle \langle 2|), \\
  \Sigma & =  \sqrt{\gamma_1} \sigma_1 + e^{i\phi} \sqrt{\gamma_2}\sigma_2 , 
\end{align}
with $\omega_0$ the frequency of both transitions, $\mathcal{D}$ is a superoperator defined by 
\begin{align}
       \mathcal{D}[c]\rho \equiv \smallfrac{1}{2} \left( 2 c \rho c^\dagger - c^\dagger c \rho - \rho c^\dagger c \right)  
    \label{superD} 
\end{align}
for an arbitrary operator $c$, and $H_{\ms{L}}$ is the Lamb shift Hamiltonian, given by 
\begin{align}
   H_{\ms{L}} & = - \hbar \biggl[ \sum_j \Delta_j \sigma_j^\dagger \sigma_j -  \sqrt{\Delta_1\Delta_2} ( e^{i\phi} \sigma_1^\dagger \sigma_2 + \mbox{H.c.} ) \biggr]  
   \label{HLdeg}
\end{align} 
with $\sigma_1 = |0\rangle \langle 1|$, $\sigma_{2} =  |0\rangle \langle 2|$. The phase $\phi$ is determined by the phases of the interactions between the transitions and the bath (see Eq.(\ref{Adef})). We also note that $H_{\ms{L}}$ can be factored as 
 $H_L = -\hbar f(\omega) D^\dagger D$ in which $D  =  \sqrt{\Delta_1} \sigma_1 + e^{i\phi} \sqrt{\Delta_2}\sigma_2 $. 

From the derivation of the degenerate master equation~\cite{Breuer07} (see also  Section~\ref{derivME}) we know that the decay rates $\gamma_j$ depend on the spectral density of the bath evaluated at their corresponding transition frequencies $\omega_j$. The Lamb shifts depend both on the damping rates (to which they are proportional) as well as a factor that is an integral of the entire spectral density. Thus if the frequencies of the transitions are changed while leaving the spectral density the same, the decay rates and Lamb shifts also change.


We find that the backed-out model for $\Delta\omega \not= 0$, in which the Hamiltonian is now   
\begin{align}
 H_0 & = \hbar \omega_0 |1\rangle \langle 1| + \hbar (\omega_0+\Delta\omega)|2\rangle \langle 2|, 
\end{align}
has exactly the same form as the degenerate master equation. That is, it can be written as Eq.(\ref{degME}) with $H_{\ms{L}} =  \sum_j \zeta_j \sigma_j^\dagger \sigma_j + ( \xi \sigma_1^\dagger \sigma_2 + \mbox{H.c.} )$ and $\sigma = \sum_j \eta_j \sigma_j$ for some set of $\{\zeta_j,\xi,\eta_j\}$. This may be considered a little surprising, given that the non-degenerate master equation has no terms in the Lamb shift Hamiltonian proportional to $\sigma_1\sigma_2$. A simple guess for the parameters $\zeta_j$, $\xi$, and $\eta_j$ as functions of the Lamb shifts and the decay rates is to take exactly the expression for the degenerate master equation, but to replace $\Delta_2(\omega_0)$ and $\gamma_2(\omega_0)$ by new values implied by the new value of $\omega_2$, namely $\Delta_2(\omega_0 + \Delta\omega)$ and $\gamma_2(\omega_0 + \Delta\omega)$. 

\begin{figure}[t] 
\centering
\leavevmode\includegraphics[width=1\hsize]{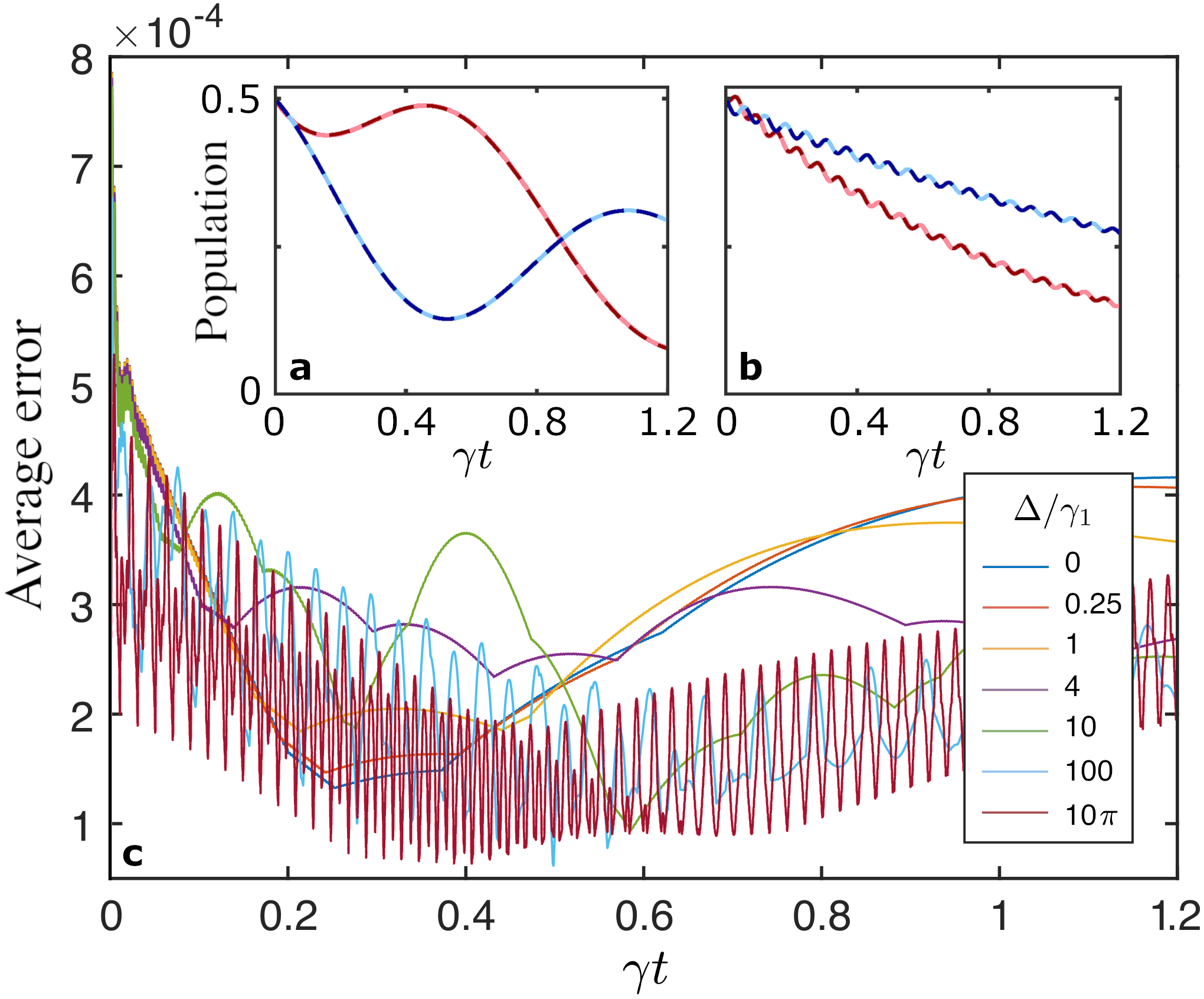}
\caption{(Color online) Here we compare the exact evolution of the weakly-damped V system with that predicted by the master equation in Eq.(\ref{degME}) with the replacements $\Delta_2(\omega_0) \rightarrow \Delta_2(\omega_0 + \Delta\omega)$ and $\gamma_2(\omega_0) \rightarrow \gamma_2(\omega_0 + \Delta\omega)$, which for the V-system is equivalent to the new Lindblad master equation we derive in Section~\ref{derivME}. In (a) and (b) we show the populations of levels $|1\rangle$ (red) and $|2\rangle$ (blue) as a function of time with the initial state $|\psi_0\rangle = (|1\rangle + |2\rangle)/\sqrt{2}$. The damping rates are $\gamma_1 = 2\gamma_2 = 0.1\tilde{\nu}$, in which $\tilde{\nu}$ is an arbitrary frequency specifying our frequency units. The solid curves are the exact evolution and the dashed curves are that of the master equation. The detuning is (a)  $\Delta\omega = 4\gamma_1$, (b) $\Delta\omega = 100\gamma_1$. In (c), for the values of the detuning shown in the legend, and with the parameters above, we plot a measure of the deviation of the master equation from the exact dynamics as a function of time. This measure is an average taken over the deviations of the four relevant elements of the density matrix: the populations of levels $|1\rangle$ and $|2\rangle$, and the real and imaginary part of the coherence between them. Notably, this average error remains less than $8\times 10^{-4}$ for all the values of detuning we explored.
}
\label{figVer}
\end{figure}

We find that this trial master equation does indeed match the model backed out using SID for all three values of $\Delta \omega$. We compare further the evolution predicted by this master equation to the exact evolution in Fig.~\ref{figVer}, for a range of values of $\Delta \omega$. For these simulations we use $\omega_0 = 10\pi$ and $\gamma_1(\omega_0) = 2 \gamma_2(\omega_0) = 0.1$. In Fig.\ref{figVer}c we show a measure of the difference between the evolution given by the master equation and exact simulations for a range of values of the detuning. This measure is an average of the absolute values of the differences between the populations and coherences of the density matrix averaged over time. The measure is below $10^{-3}$ for all values of the detuning shown. 

In Fig.\ref{figVer}b we plot the evolution of the populations of the upper levels for both the master equation and the exact simulation for a value of $\Delta\omega$ that might be considered well into the non-degenerate regime ($\Delta\omega = 100\gamma$). We see that the evolution contains ``wiggles'' that are correctly predicted by our trial master equation, but are not predicted by the non-degenerate master equation. These wiggles vanish as $\Delta\omega \rightarrow \infty$. 

In showing that, for the V-system, the degenerate, near-degenerate, and far-detuned regimes are \textit{all} described by a Lindblad master equation that is essentially the degenerate master equation, SID has provided us with the insight we need to derive this Lindblad equation from the B-R equation. An inspection of the B-R equation, Eq.(\ref{BR}) in the next section, shows that the only way that it can take this Lindblad form is if the frequency-dependent parameters $R_j$ and $I_j$ that appear in this equation are effectively equal for transitions whose frequency differences are on the order of the damping rates and Lamb shifts, respectively. Combining this with the fact that certain terms containing these parameters are effectively removed by the rotating-wave approximation when the same frequency differences are much larger than the damping rates and Lamb shifts allows us to derive the Lindblad master equation. We perform this derivation in the next section. 

\section{Derivation of the master equation}
\label{derivME}
 
Having shown numerically that there is a Markovian master equation describing two arbitrarily detuned transitions, at least for the Ohmic bath, as well as obtaining the form that this equation takes, a close examination of the usual derivation of the existing Markovian master equations reveals how this more general master equation can be derived for an arbitrary number of levels. For simplicity we present this derivation first for a bath at zero temperature. We then outline the derivation for non-zero temperature, since it is essentially the same.   

The Hamiltonian for the system and the bath, in which the latter consists of a continuum of independent harmonic oscillators, is given by 
\begin{align} 
    H & = H_{\ms{sys}} + \hbar (A + A^\dagger)\int_0^\Omega \!\!\!\! \sqrt{J(\omega)} \left[ b(\omega) + b^\dagger(\omega)\right] d\omega \nn \\
    & \;\;\; + \int_0^\Omega \!\! \hbar \omega b(\omega)^\dagger b(\omega) \, d\omega . 
    \label{sysbath}
\end{align} 
Here $H_{\ms{sys}}$ is the Hamiltonian of the system which has a discrete set of energy levels. The operators $b(\omega)$ are the annihilation operators for a continuum of harmonic oscillators indexed by their frequencies $\omega$ (equivalently the modes of a quantum field). The function $J(\omega)$ is the density of oscillators per unit frequency, usually referred to as the \textit{spectral density} of the bath. The maximum frequency of the bath oscillators is $\Omega$ and is called the ``cut-off'' frequency. Instead of having sharp cut-off at frequency $\Omega$ one can instead arrange $J(\omega)$ to exhibit a smooth drop-off above some frequency. We use a sharp cut-off purely for simplicity. As will be clear in what follows, in the weak-damping regime the addition of a smooth cut-off merely modifies the values of the damping rates and Lamb shifts.  

We have written the Hermitian operator of the system that couples to the bath  as $A+A^\dagger$. Here $A$ is defined as containing all the matrix elements of this Hermitian operator that transform higher energy levels to lower ones. If we denote the energy levels of the system by $|n\rangle$, thus writing
\begin{align}
    H_{\ms{sys}} = \sum_n E_n |n\rangle \langle n | , 
\end{align}
and define transition (or ``decay'') operators 
\begin{align}
    \sigma_j = |n_j\rangle \langle m_j| , \;\; E_{m_j}  > E_{n_j}, \;\; j = 1,\ldots, J , 
\end{align}
then $A$ can be written as   
\begin{align}
    A & =  \sum_{j=1}^J g_j \sigma_j   ,  \label{Adef}
\end{align}
where $g_j = |g_j|e^{i\phi_j}$ are complex numbers giving the magnitude and phase of the coupling of transition $j$ to the bath. The frequency of transition $j$ is  
\begin{align}
    \omega_j = \frac{E_{m_j} - E_{n_j}}{\hbar} , 
\end{align}
and the evolution of the transition operators $\sigma_j$ and bath operators $b(\omega)$ in the interaction picture is
\begin{align} 
   \sigma_j^{\ms{I}} & = \sigma_j e^{-i\omega_j t} , \\ 
   b^{\ms{I}}(\omega) & = b(\omega) e^{-i\omega t} . 
\end{align} 

To derive Markovian master equations one applies a rotating-wave approximation (RWA) to the Hamiltonian above. This should not be confused with a second rotating-wave approximation which is the final step that turns the Bloch-Redfeild master equation into the degenerate and non-degenerate Lindblad master equations. To apply the first RWA we move into the interaction picture. The terms in the interaction Hamiltonian that contain the products $Ab(\omega)$ and $A^\dagger b^\dagger(\omega)$ (the so-called ``off-resonant'' terms) become  
\begin{align} 
    H_{\ms{OR}}^{\ms{I}} & = \hbar\int_0^\Omega \!\!\!\! \sqrt{J(\omega)} \Biggl[ \sum_j g_j \sigma_j b(\omega) e^{-i(\omega_j + \omega)} + \mbox{H.c}\Biggr] d\omega . 
    \label{HOR}
\end{align} 
Since the minimum frequency at which each of the terms in the sum over $j$ oscillates is $\omega_j$, when the damping rates and Lamb shifts (to be derived below) are much less than all the $\omega_j$, these terms will average to zero on the timescale of the dynamics induced by the bath, and can be discarded. With this approximation the Hamiltonian of the system and bath becomes 
\begin{align} 
    H_{\ms{RWA}} & = H_{\ms{sys}} + \hbar\int_0^\Omega \!\!\!\! \sqrt{J(\omega)} \left[ A^\dagger b(\omega) + A b^\dagger(\omega)\right] d\omega \nn \\
    & \;\;\; + \int_0^\Omega \!\! \hbar \omega b(\omega)^\dagger b(\omega) \, d\omega . 
    \label{sysbathRWA} 
\end{align} 
The regime of weak damping is defined as the regime in which we are close enough to the limit in which $\min_j(\omega_j)/\max_j(\gamma_j) \rightarrow \infty$ so that this approximation is a good one.  

To proceed one now applies what are known as the Born-Markov approximations to the evolution generated by $H_{\ms{RWA}}$. For the details of these approximations we refer the reader to~\cite{WM10, Breuer07, Gardiner10}). The result is the following expression for the evolution of the density matrix of the system in the interaction picture: 
\begin{align}
    \frac{d\rho^{\ms{I}}}{dt}  & = -\frac{1}{\hbar^2} \int_0^\infty \!\!\! \mbox{Tr}_{\ms{B}} \left[ H^{\ms{I}}_{\ms{R}}(t) , \left[ H^{\ms{I}}_{\ms{R}}(s), \rho^{\ms{I}}(t)\otimes \rho_{\ms{B}}(0) \right] \right] ds ,
\end{align}
where 
\begin{align}
    H_{\ms{R}}^{\ms{I}} = \hbar\int_0^\Omega \!\!\!\! \sqrt{J(\omega)} \Biggl[  \sum_j g_j \sigma_j^{\ms{I}\dagger} b^{\ms{I}}(\omega) + \mbox{H.c}\Biggr] d\omega 
\end{align}  
is the interaction between the system and the bath that appears in $H_{\ms{RWA}}$, in the interaction picture. The operator $\rho^{\ms{I}}(t)$ is the density matrix of the system in the interaction picture, $\rho_{\ms{B}}(0)$ is the initial density matrix of the bath, and $\mbox{Tr}_{\ms{B}}[\cdot]$ denotes the trace over the bath. 

To proceed now we will examine a single term from the expression above, since all the terms are similar and each is processed in the same way. Substituting $H_{\ms{R}}^{\ms{I}}(t)$ and $H_{\ms{R}}^{\ms{I}}(s)$ into the expression above, one of the terms we obtain is  
\begin{align} 
   K & = \int_0^\infty  \left[ \int_0^\Omega \!\! G(\omega) \, d\omega  \right] \sigma_k^{\ms{I}\dagger}(t) \sigma_j^{\ms{I}}(s) \rho^{\ms{I}}(t) \, ds , 
\end{align}
where 
\begin{align}
    G(\omega) & = e^{-i\omega (t-s)} \!\! \int_0^\Omega \!  \left\langle b(\omega) b^\dagger(\omega')  \right\rangle \sqrt{J(\omega)J(\omega')} \,  d\omega' \nonumber \\ 
       & = J(\omega) e^{-i\omega (t-s)}  . 
\end{align}
Here we have used the relation  $[b(\omega),b^\dagger (\omega')] = \delta(\omega-\omega') $
and chosen the field to be at zero temperature so that $\left\langle b^\dagger(\omega')b(\omega)\right\rangle = 0$.

Now substituting $G(\omega)$ into $K$ and rearranging we obtain 
\begin{align} 
K & = \int_0^\infty  \left[ \int_0^\Omega \!\! G(\omega)\, d\omega \right] \sigma_k^{\ms{I}\dagger}(t) \sigma_j^{\ms{I}}(s) \rho^{\ms{I}}(t) ds \nonumber \\ 
   & = \left[ \int_0^\infty \!\! \int_0^\Omega \!\! G(\omega-\omega_j)\, d\omega \, ds \right] e^{i(\omega_k - \omega_j) t} \sigma_k^{\dagger} \sigma_j   \rho^{\ms{I}}(t) \nonumber \\ 
      & = \left[ \int_0^\Omega \!\! \int_0^\infty \!\! G(\omega-\omega_j)\, ds \, d\omega \right]  \sigma_k^{\ms{I}\dagger}(t) \sigma_j^{\ms{I}}(t)   \rho^{\ms{I}}(t) . 
   \label{meterm}
\end{align}
It is useful to define  
\begin{align}
    \Gamma_j & \equiv \Gamma(\omega_j) \equiv  \int_0^\Omega \!\! \int_0^\infty \!\! G(\omega-\omega_j)\, ds \, d\omega , \\
    R_j & = \mbox{Re}[\Gamma_j] , \\
    I_j & = \mbox{Im}[\Gamma_j] . 
\end{align} 
Moving back into the Schr\"{o}dinger picture, and writing down all the terms, we obtain the Bloch-Redfield equation, which is 
\begin{align}
    \dot\rho & = -\frac{i}{\hbar} [H_{\ms{sys}},\rho]  - i\sum_j I_j [\Sigma_j^\dagger \Sigma_j,\rho] + \sum_j 2 R_j \mathcal{D}[\Sigma_j]\rho \nonumber \\
    & \;\;\; \; - i \sum_{k\not=j} I_k \left[  \Sigma_j^\dagger \Sigma_k \rho - \rho \Sigma_k^\dagger \Sigma_j +  \Sigma_j \rho \Sigma_k^\dagger  - \Sigma_k \rho \Sigma_j^\dagger  \right]  \nonumber \\ 
    & \;\;\;\; - \sum_{k\not=j} R_k   \left[ \Sigma_j^\dagger \Sigma_k \rho +  \rho \Sigma_k^\dagger \Sigma_j - \Sigma_j \rho \Sigma_k^\dagger -  \Sigma_k \rho \Sigma_j^\dagger \right] . 
    \label{BR}
\end{align}
Here, for compactness, we have defined 
\begin{align}
    \Sigma_j \equiv g_j \sigma_j , 
\end{align}
and $\mathcal{D}$ is the superoperator defined in Eq.(\ref{superD}). 

We note that the decay rates $\gamma_j$ will be 
\begin{align}
    \gamma_j &= 2 |g_j|^2 R_j \nonumber \\ 
    & =  2 |g_j|^2 \int_0^\Omega \!\!\! J(\omega) \left[ \int_0^\infty \!\!\! \cos([\omega - \omega_j]s) ds\right] d\omega \nn \\ 
    & = 2 |g_j|^2 \int_0^\Omega \!\!\! J(\omega) \pi \delta(\omega - \omega_j) d\omega \nn \\ 
    & = 2 \pi |g_j|^2 J(\omega_j) \nn \\
    & \equiv \gamma (\omega_j)   \label{gammadef}
\end{align} 
and the Lamb shifts will be  
\begin{align} 
   \Delta_j & = |g_j|^2 \, I_j  \nonumber \\ 
   & =  |g_j|^2 \int_0^\Omega \!\!\! J(\omega) \left[ \int_0^\infty \!\!\! \sin([\omega - \omega_j]s) ds\right] d\omega \nn \\ 
    & =  |g_j|^2 \, \mathbb{P}\left[ \int_0^\Omega \!\!\! J(\omega) \left( \frac{1}{\omega - \omega_j} \right) d\omega \right] \nn \\ 
    & = |g_j|^2 \, \mathbb{P} \left[ \int_{-\omega_j}^{\Omega-\omega_j}   \frac{J(\omega + \omega_j) }{\omega}  d\omega \right] \nn \\
    & \equiv \Delta (\omega_j) .
\end{align}
Here $\mathbb{P}[\cdot]$ denotes the \textit{principle value} of an integral. For readers not familiar with this quantity we give the definition and an example in Appendix~\ref{PV}. So long as the spectral density does not decrease with $\omega$, and $\Omega > \omega_j$, the Lamb shift $\Delta_j$ can be expected to be greater than the damping rate $\gamma_j$ (this is true for the Ohmic bath, see below).   
 
The master equation we have derived in Eq.(\ref{BR}), the Bloch-Redfield equation, includes arbitrary detuning between the levels, but it is not in the Lindblad form, and does not guarantee that the density matrix will remain positive. We wish to obtain a master equation in the Lindblad form that is still valid for all detunings between the transitions.  

Note first that when transitions $j$ and $k$ are degenerate $\Gamma_j = \Gamma_k$. In this case the last two lines of Eq.(\ref{BR}) combine respectively with the last two terms on the first line to give the degenerate master equation for these transitions. In this case the Lamb-shift Hamiltonian is $H_{\ms{L}}= \hbar D^\dagger D $ with $D = \sqrt{I_j}\Sigma_j+\sqrt{I_k}\Sigma_k$ and the Lindblad damping term is $\mathcal{D}[\sqrt{\gamma_j}\Sigma_j + \sqrt{\gamma_k}\Sigma_k]$.

There is another situation in which the B-R equation will reduce immediately to a Lindblad master equation: if the spectrum is flat, meaning that it is the same for all $\omega$. (The assumption of a flat spectrum is often called the ``white noise approximation", as it is useful for deriving quantum Langevin equations for open systems in the non-degenerate regime~\cite{Collett84, Gardiner85}.) If the spectrum is flat then $\Gamma_j = \Gamma_k$ for all values of $\omega_j$ and $\omega_k$, and so in this case the resulting master equation, which has the same form as the degenerate master equation, is valid for all regimes. Unfortunately, physically relevant spectra are not flat. 

What we do now is to show that a flat spectrum is not required to derive a master equation valid for all regimes; the rate of change of the spectrum with respect to $\omega$ does not need to be zero, it merely needs to be small \textit{enough}. We determine the necessary condition on this rate of change, and use it to derive the Lindblad master equation. This master equation has the same form as that for a flat spectrum, but of course the damping rates and Lamb shifts will not be the same as those for a flat spectrum, since the spectral density will now in general be different for different transitions. 

To begin we observe that when two transition frequencies, $\omega_j$ and $\omega_k$, are different, every term in the last two lines of Eq.(\ref{BR}) will oscillate at the difference frequency $\Delta\omega_{jk} = |\omega_{j}-\omega_{k}|$. If this difference frequency is sufficiently high then these terms will average to zero and we will be left only with the first line of Eq.(\ref{BR}), which is the non-degenerate master equation. How large does $\Delta\omega_{jk}$ need to be to eliminate the last two lines of Eq.(\ref{BR})? It needs to be much larger than the magnitudes of the rest of the dynamical terms in the master equation (excluding the Hamiltonian of the system, since this does not change the populations of the system's eigenstates). The magnitudes of the second and third terms on the top line are $\Delta_j$ and $\gamma_j$, respectively, and those on the last two lines are 
\begin{align} 
    M_{jk} & \equiv |g_jg_k| R_j \sim \sqrt{\gamma_j \gamma_k} ,    \\
    O_{jk} & \equiv  |g_jg_k| I_j \sim \sqrt{\Delta_j\Delta_k} .  
\end{align} 
So the terms on the last two lines are eliminated when $\min_{jk} \Delta\omega_{jk} \gg \max_{lm} M_{lm}$ and $\min_{jk} \Delta\omega_{jk} \gg \max_{lm} O_{lm}$. Without loss of generality we will assume that the Lamb shifts are greater than the damping rates (in the opposite case one merely switches the roles of $O_j$ and $M_j$). To find a set of terms that are in the Lindblad form, and that are an excellent approximation to the last two lines of Eq.(\ref{BR}), we only need concern ourselves with the regime 
\begin{align} 
      \Delta\omega_{jk} & \lesssim \min_{jk} O_{jk} 
      \label{reg1}
\end{align}
(since we have assumed $O_{jk} \geq M_{jk}$, the regime $\Delta\omega_{jk} \lesssim \min_{jk} O_{jk}$ automatically includes the regime $\Delta\omega_{jk} \lesssim \min_{jk} M_{jk}$). Outside of this regime, the last two lines will be eliminated by the oscillations at the detuning frequency $\Delta\omega_{jk}$. 

We now recall that $\gamma_j$ and $\Delta_j$, and therefore $M_{jk}$ and $O_{jk}$, must be much smaller than both transition frequencies in order for the master equation to be valid. This is a requirement of the initial rotating wave approximation discussed above. Combining this with Eq.(\ref{reg1}) we need only consider the regime in which 
\begin{align}
    \Delta\omega_{jk} \lesssim O_{jk} \sim \sqrt{\Delta_j\Delta_k} \ll \sqrt{\omega_j \omega_k} .   
\end{align}
Now if $\Gamma_j$ (and thus $R_j$ and $I_j$) does not vary rapidly on the scale of $O_{jk} \sim \sqrt{\Delta_j \Delta_k}$ (which is the scale of the Lamb shifts), then in the regime we need to consider we have $\Gamma_j \approx \Gamma_k$. More specifically, we consider systems for which the spectral density satisfies
\begin{align}
     J(\omega_j + \Delta_k) \approx  J(\omega_j), \;\;\; \forall j,k, 
\end{align}
for which the more precise statement is  
\begin{align}
     |J(\omega_j + \Delta_k) - J(\omega_j)| \ll  J(\omega_j), \;\;\; \forall j,k,   \label{Jslow}
\end{align}
since this implies that $\Gamma(\omega)$ also satisfies the same ``slow variation'' conditions. Under the condition in Eq.(\ref{Jslow}) we have 
\begin{align}
   |\omega_j - \omega_k| & \lesssim O_{jk} \;\; \Rightarrow \;\; \left\{ 
   \begin{array}{l}
       R_{j}  \approx R_k   
       \\
       I_{j} \approx I_k
   \end{array}  \right. \label{Rsim} 
\end{align}
while at the same time allowing 
\begin{align}
  R_{j} & \not=  R_k, \; I_{j} \not=  I_k  \;\; \mbox{when} \;\; |\omega_j - \omega_k| \gg O_{jk} , 
\end{align}
The relation (\ref{Rsim}) allows us to make the replacements $R_{j} \approx \sqrt{R_j R_k}$ and $I_{j} \approx \sqrt{I_j I_k}$ in Eq.(\ref{BR}) because i) when $|\omega_j - \omega_k| \lesssim O_{jk}$ these replacements are well justified, and ii) when $|\omega_j - \omega_k|$ is larger the terms containing $\sqrt{R_j R_k}$ and $\sqrt{I_j I_k}$ are eliminated by the rotating wave approximation. The result is 
\begin{align}
    \dot\rho & = -\frac{i}{\hbar} [H_{\ms{sys}},\rho]  - i\sum_j I_j [\Sigma_j^\dagger \Sigma_j,\rho] + \sum_j 2 R_j \mathcal{D}[\Sigma_j]\rho \nonumber \\
    & \;\;\; \; - i \sum_{k\not=j} \sqrt{I_j I_k} \left[  \Sigma_j^\dagger \Sigma_k + \Sigma_k^\dagger \Sigma_j, \rho \right]  \nonumber \\ 
    & \;\;\;\; - 2 \sum_{k\not=j} \sqrt{R_j R_k}   \left[ \Sigma_j^\dagger \Sigma_k \rho +  \rho \Sigma_k^\dagger \Sigma_j - 2 \Sigma_j \rho \Sigma_k^\dagger  \right] . 
    \label{me1}
\end{align}
The terms in this equation can be re-factored so as to write it in a much neater form, namely 
\begin{align} 
   \dot\rho & = -\frac{i}{\hbar}\left[H_0 - \hbar  D^\dagger D,\rho \right] - \mathcal{D}[\Sigma] \rho  \label{HQM} 
\end{align} 
This is the zero-temperature Lindblad-form master equation for all regimes. The operators $\Sigma$ and $D$ are 
\begin{align}
        \Sigma & = \sum_{j=1}^N \sqrt{\gamma_j} e^{i\phi_j} \sigma_j, \\
        D  & = \sum_{j=1}^N \sqrt{\Delta_j} e^{i\phi_j} \sigma_j , 
\end{align} 
in which $\phi_j = \arg[g_j]$ as defined below Eq.(\ref{Adef}). The term $H_{\ms{L}} \equiv -\hbar D^\dagger D$ is the Lamb-shift Hamiltonian. If we expand it out we see that, so long as $\sigma_j^\dagger \sigma_k \not=0$, the upper levels of the transitions $j$ and $k$ are coupled together via the bath: 
\begin{align} 
    H_{\ms{L}}  & =   -\hbar \sum_{j=1}^N  \Delta_j  \sigma_j^\dagger \sigma_j  \nn \\
    & \;\;\;\; -\!\hbar \sum_{j=1}^N\sum_{k=j+1}^N \sqrt{ \Delta_j  \Delta_k } \left( e^{i\Delta\phi_{jk}} \sigma_j^\dagger \sigma_k + \mbox{H.c.} \right) ,   
\end{align} 
where the phases are given by 
\begin{align}
    \Delta\phi_{jk} & = \phi_k - \phi_j .   
\end{align} 
The decay rates $\gamma_j$ are given in Eq.(\ref{gammadef}) and are determined by the value of the spectral density $J(\omega)$ only at transition frequency $\omega_j$. The Lamb shifts, on the other hand, depend on the whole spectral density, and in particular on the cut-off frequency. As an example, for the Ohmic spectrum with a sharp cut-off at $\Omega$, in which $J(\omega) \propto \omega$ (we choose to define $J(\omega) = \omega/\Omega^2$) the Lamb shifts are \begin{align} 
   \Delta_j & = |g_j|^2 \, \mathbb{P} \left[ \int_{-\omega_j}^{\Omega-\omega_j}   \frac{J(\omega + \omega_j) }{\omega}  d\omega \right] \nn \\
     & = \frac{|g_j|^2}{\Omega^2} \, \mathbb{P} \left[ \int_{-\omega_j}^{\Omega-\omega_j}   \frac{(\omega + \omega_j) }{\omega}  d\omega \right] \nn \\
     & = \frac{\gamma_j}{2\pi} \left[ \frac{\Omega}{\omega_j} +  \ln\left( \frac{\Omega}{\omega_j} - 1\right)  \right] . 
     \label{eq42}
\end{align} 
We see that so long as $\Omega$ is larger than $2\omega_j$ the Lamb shift is larger than the damping rate by at least a factor of $\Omega/\omega_j$. Recall that the Lamb shifts are  required to be much less that the transition frequencies. Using the expression for $\Delta_j$ above, we have 
\begin{align} 
   \frac{\Delta_j}{\omega_j} 
     & = \frac{\gamma_j}{2\pi} \left[ \frac{\Omega}{\omega_j} +  \ln\left( \frac{\Omega}{\omega_j} - 1\right)  \right] \nn \\
     & \approx \frac{1}{2\pi} \left(\frac{\gamma_j}{\omega_j} \right) \left(\frac{\Omega}{\omega_j} \right)  . 
\end{align} 
Thus to satisfy the condition $\Delta_j \ll \omega_j$  requires that the cut-off frequency is not too large. In particular $\Omega \ll 2\pi \omega_j^2/\gamma_j$. 

We can now evaluate the fidelity of our approximation explicitly for the Ohmic spectrum. Recall that we require $\Delta_k/\Delta_j \approx 1$, when $\Delta\omega = |\omega_j - \omega_k| \lesssim  \Delta_j$. Denoting $\Delta_j$ by  $\Delta(\omega_j)$, and writing $\omega_k = \omega_j + \Delta_j$, we have 
\begin{align} 
   \frac{\Delta(\omega_j +\Delta_j)}{\Delta(\omega_j)} - 1 & = \frac{\Delta_j}{\Omega} \ln\left( \frac{\Omega}{\omega_j} - 1\right) + \mathcal{O}(\Delta\omega^2) \nn \\
   & \approx \left[ \frac{\Delta_j}{\omega_j} \right] \left[ \frac{\omega_j}{\Omega} \ln\left( \frac{\Omega}{\omega_j} \right) \right] \ll 1 . 
\end{align} 
Since the master equation is already derived under the conditions that $\Omega \gg \omega_j$, and $\omega_j \gg \Delta_j$, the expressions in \textit{both} of the square brackets are individually much less than unity. Thus the slowly varying spectrum approximation that we have introduced is automatically a very good approximation for the Ohmic bath. 

\subsection*{Master equation for arbitrary temperature} 
\label{nonz}

To derive the master equation for non-zero temperature we merely replace the zero-temperature expectation values of the bath operators with their expectation values at non-zero temperature, which are 
\begin{align}
    \left\langle b^\dagger(\omega')b(\omega)\right\rangle & = n_{T}(\omega) \delta(\omega-\omega') , \\
    \left\langle b(\omega')b^\dagger(\omega)\right\rangle & = [1 + n_{T}(\omega)] \delta(\omega-\omega') , 
\end{align} 
in which 
\begin{align}
    n_{T}(\omega) & =  \frac{1}{\exp\left[\hbar\omega/\left(k_{\ms{B}}T\right)\right] -1}  . 
\end{align} 
Here $T$ is the temperature of the bath and $k_{\ms{B}}$ is Boltzmann's constant. With these new expectation values we now obtain more terms in the master equation. For the new terms the spectral density $J(\omega)$ is multiplied by $n_T(\omega)$, so the new terms give a new integral for which we have to calculate the principle value. This integral is 
\begin{align}
    \mbox{Im}[\tilde{\Gamma}_j^{\ms{T}}] & =  \mathbb{P} \left[ \int_{-\omega_j}^{\Omega-\omega_j}   \frac{J(\omega + \omega_j)n_T(\omega + \omega_j) }{\omega}  d\omega \right] \nn \\
    & = \frac{1}{\Omega^2}\mathbb{P} \left[ \int_{-\omega_j}^{\Omega-\omega_j}  \!\! \frac{1 + (\omega_j/\omega) }{\exp[\hbar(\omega+\omega_j)/(k_{\ms{B}}T)]-1}  d\omega \right] . 
\end{align} 
Unfortunately this integral does not have an analytic solution, so we leave it as an integral and define a new set of Lamb shifts   
\begin{align} 
    \Delta_j^{\ms{T}} \equiv |g_j|^2\mbox{Im}[\tilde{\Gamma}_j^{\ms{T}}] . 
\end{align} 
The approximations we used for the zero temperature part of the master equation can be applied in exactly the same way to the new terms that appear at non-zero temperature. The resulting master equation for arbitrary temperatures is 
\begin{align} 
   \dot\rho & = -\frac{i}{\hbar}[H_0+H_{\ms{L}},\rho] - \mathcal{D}[\Theta(T)] \rho - \mathcal{D}[\Upsilon(T)] \rho ,  \label{HQMT} 
\end{align} 
where 
\begin{align} 
       \Theta(T) & = \sum_{j=1}^N \sqrt{\gamma_j\left[1 + n_T(\omega_j)\right]} \, e^{i\phi_j} \sigma_j , \\
       \Upsilon(T) & = \sum_{j=1}^N  \sqrt{\gamma_j n_T(\omega_j)} \, e^{-i\phi_j} \sigma_j^\dagger ,
\end{align} 
and
\begin{align}
    H_{\ms{L}}  & = -\hbar \left[  B^\dagger B - C C^\dagger \right] , 
\end{align}
with  
\begin{align}
    B & = \sum_{j=1}^N \sqrt{ \Delta_j + \Delta_j^{\ms{T}}} \, e^{i\phi_j} \sigma_j,\\ 
    C & =  \sum_{j=1}^N \sqrt{ \Delta_j^{\ms{T}} } \, e^{i\phi_j} \sigma_j . 
\end{align}
Note that when $T>0$ the bath induces a Hamiltonian coupling not only between the upper levels of the different transitions, but also the lower levels. 

\subsection*{The non-degenerate master equation, the secular approximation, and numerical efficiency}

In deriving the master equation valid for all regimes, we did not make the secular approximation, which involves dropping terms that oscillate at the frequency difference between different transitions. However, when the difference between the frequencies of two transitions is much larger than the Lamb shifts and linewidths, keeping the resulting rapidly oscillating terms greatly increases the numerical overhead while contributing little to the evolution. In this case one should drop these  terms for numerical efficiency. Doing so transforms the master equation into the non-degenerate master equation, but only for those pairs of transitions for which the detuning is very large. If we write the master equation in the form given in Eq.(\ref{me1}), then dropping the rapidly oscillating terms means merely dropping terms in the second and third lines for the pairs of values of $j$ and $k$ whose transitions are detuned by much more that their Lamb shifts and linewidths. 

For readers very familiar with the degenerate and non-degenerate master equations, the result of applying the secular approximation to a subset of pairs of transitions will likely be clear. For readers without this familiarity, we give an explicit example. Let us say that we can divide our transitions into two sets, where the frequencies of those in the first set differ from the frequencies of those in the second set by at least $10^3\Delta_{\ms{max}}$ in which $\Delta_{\ms{max}}$ is the maximum Lamb shift among all the transitions. If we denote the transition operators in the first set by $\sigma^{(1)}_j$, with $j = 1,\ldots,N_1$, and those in the second by $\sigma^{(2)}_j$, with $j = 1,\ldots,N_2$, then the result of making the secular approximation on the master equation in Eq.(\ref{HQM}) is 
\begin{align} 
   \dot\rho & = -\frac{i}{\hbar}\left[H_0 - \hbar  \sum_{m=1}^2 D^\dagger_m D_m,\rho \right] - \sum_{m=1}^2 \mathcal{D}[\Sigma_m] \rho, \label{secular} 
\end{align} 
with   
\begin{align}
        \Sigma_m & = \sum_{j=1}^{N_m} \sqrt{\gamma_j^{(m)}} e^{i\phi_j^{(m)}} \sigma_j^{(m)}, \\
        D_m  & = \sum_{j=1}^{N_m} \sqrt{\Delta_j^{(m)}} e^{i\phi_j^{(m)}} \sigma_j^{(m)} .  
\end{align} 

\section{Accuracy of the Lindblad Master Equation: additional confirmation for the Ohmic bath}
\label{accnum}

It is clear from the derivation in the previous section that the master equation we have obtained, given in Eqs.(\ref{HQM}) and (\ref{HQMT}), will be valid so long as the variation of the Lamb shifts $\Delta (\omega_j)$ and damping rates $\gamma(\omega_j)$ on the scale of these same Lamb shifts and damping rates is sufficiently small. This variation of the Lamb shifts and damping rates will be small if the variation of the spectral density $J(\omega)$ on the scale of the Lamb shifts and damping rates is sufficiently small. Below we will verify quantitatively the accuracy of the new master equation for the Ohmic spectrum at zero temperature, using exact simulations of two example systems. The Ohmic spectrum is appropriate for systems such as atoms, color centers, or superconducting qubits coupled to one-dimensional wave-guides or transmission lines.  

Since the efficacy of the approximation used to derive the master equation does not depend on the temperature of the bath or the specific functional form of the spectral density (it depends only on the local variation of the resulting Lamb shifts and damping rates around their respective transition frequencies), simulations for the Ohmic bath at zero temperature provide a high level of confidence in the accuracy of the master equation more generally. Further, since we have exact analytic expressions for the Lamb shifts and damping rates in this case, if desired the variations of these quantities at the transition frequencies can be related directly to the accuracy determined in our simulations.

\begin{figure}[t] 
\centering
\leavevmode\includegraphics[width=1\hsize]{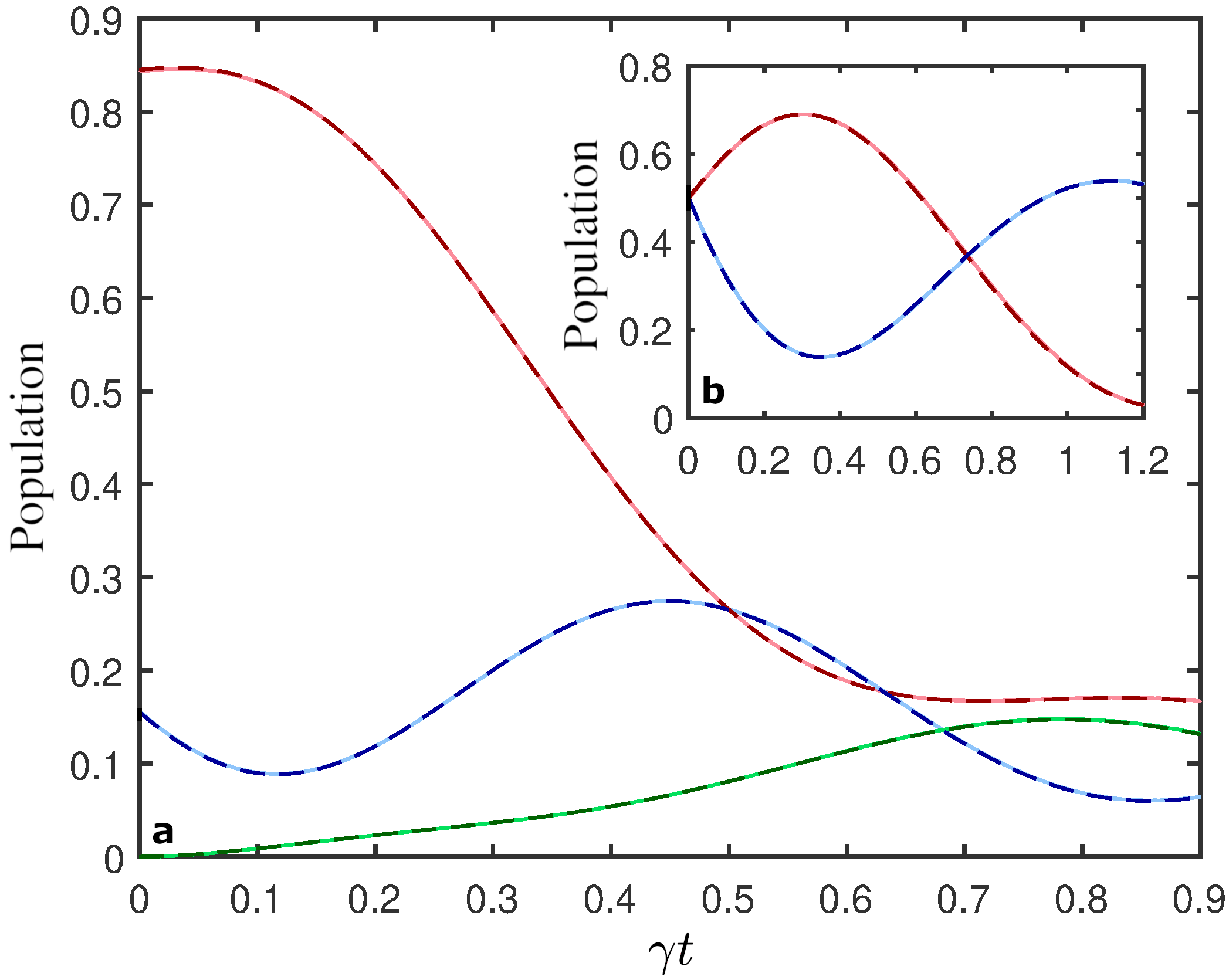}
\caption{(Color online) A comparison of the master equation given in Eq.(\ref{HQM}) with the exact evolution for two open systems. The evolution of the master equation is shown as dashed lines and the exact evolution as solid lines. (a) The populations of the three upper levels of the trident system depicted in Fig.\ref{6systems}b with initial state $|\psi_0\rangle = (7i|1\rangle + 3|2\rangle)/\sqrt{58}$. (b) The populations of the two upper levels of the two-qubit system depicted in Fig.\ref{6systems}d with the initial state $|\psi_0\rangle = (i|1\rangle_1|0\rangle_2 + |0\rangle_1|1\rangle_2)/\sqrt{2}$ and the parameters $\gamma_1 = \gamma_2 = 0.1\tilde{\nu}$, $\omega_1 = 10\pi\tilde{\nu}$, and $\omega_2 = \omega_1 + 2\gamma$.  
}
\label{figTri2Q}
\end{figure}

We have already compared the evolution of the master equation to exact simulations for the V system in Fig.~\ref{figVer}. We now consider two further systems. The first is the ``trident'' system depicted in Fig.\ref{6systems}b. This system has three transitions, and thus also three pairs of transitions which we can place simultaneously in the near-degenerate regime. Again using $\tilde{\nu}$ as our arbitrary frequency reference, we choose  parameters $\omega_1 = 10\pi\tilde{\nu}$, $\omega_j = \omega_1 + \gamma_2/(j-1)$, for $j=2,3$, and $\gamma_j = [(5-j)/40]\tilde{\nu}$, for $j=1,2,3$, with the cut-off frequency $\Omega = 80\pi\tilde{\nu}$. Choosing the initial state $|\psi_0\rangle = (7i|1\rangle + 3|2\rangle)/\sqrt{58}$, we plot the evolution of the populations predicted by the master equation along with the exact evolution in Fig.~\ref{figTri2Q}a. The maximum error in the evolution of the master equation over the time period plotted in Fig.~\ref{figTri2Q}a is less than $2\times 10^{-3}$.

We now perform simulations for two co-located qubits, whose level structure is depicted in Fig.~\ref{6systems}d. We choose the parameters $\omega_1 = 10\pi\tilde{\nu}$, and $\omega_2 = \omega_1 + 2\gamma$, $\gamma_1 = \gamma_2 = 0.1\tilde{\nu}$, with the same cut-off frequency as before. We find that this system requires significantly larger values of the weak damping parameters (``quality factors''), $Q_j \equiv \omega_j/\gamma_j$, in order for the master equation to accurately model the dynamics. Since available numerical resources place restrictions on the sizes of the $Q_j$'s that we can practically simulate, for this system we apply the \textit{first} rotating-wave approximation to our model Hamiltonian prior to performing the exact simulations. That is, we simulate the Hamiltonian $H_{\ms{RWA}}$ (Eq.(\ref{sysbathRWA})) instead of the full model in Eq.(\ref{sysbath}). These simulations thus show us how well the master equation will perform so long as the $Q_j$'s are large enough to satisfy the first rotating-wave approximation. We stress that the values of the $Q_j$'s we actually simulate here are not large enough to satisfy this approximation for this system. This fact is interesting in itself, because it shows that different systems, even with only a few levels, can require quite different quality factors to reach the weak damping regime. We believe this is due to the availability of more channels via which the off-resonant terms in the system bath interaction can excite the two-qubit system over the V and trident systems. 

Choosing the initial state $|\psi_0\rangle = (i|1\rangle_1|0\rangle_2 + |0\rangle_1|1\rangle_2)/\sqrt{2}$, we show the evolution of the  populations for both the master equation and the exact simulations of $H_{\ms{RWA}}$ in Fig.~\ref{figTri2Q}b. The error in the evolution of the master equation over the duration shown in Fig.~\ref{figTri2Q}a is less than $5\times 10^{-3}$. 

Finally, we compare the evolution of the Bloch-Redfield equation (Eq.(\ref{BR})) both to our Lindblad master equation and the exact simulations. These comparisons, plots of which are given in the supplemental material~\footnote{The supplemental material can be found at xxxxxx}, confirm that the B-R equation and our master equation have essentially the same accuracy. This result is implied, of course, by the derivation of the Lindblad master equation. 

\section{Regime of Validity}

We have shown that when the spectral density is sufficient flat the B-R equation can be replaced by a Lindblad equation, and that this is an excellent approximation for the Ohmic bath. The question we need to answer now is whether the slowly varying spectrum (SVS) approximation remains an excellent approximation over the entire domain in which the B-R equation itself is valid, or whether there is a regime in which the B-R equation is valid but the Lindblad equation is not. Since the SVS approximation depends solely on the slope of the spectral density (strictly, the difference between the values of the spectral density at the frequencies of nearby transitions) the question is, as we increase this slope, does the B-R equation deviate from the exact evolution before or after the Lindblad and B-R equations deviate from each other? 

\begin{figure}[t] 
\centering
\leavevmode\includegraphics[width=1\hsize]{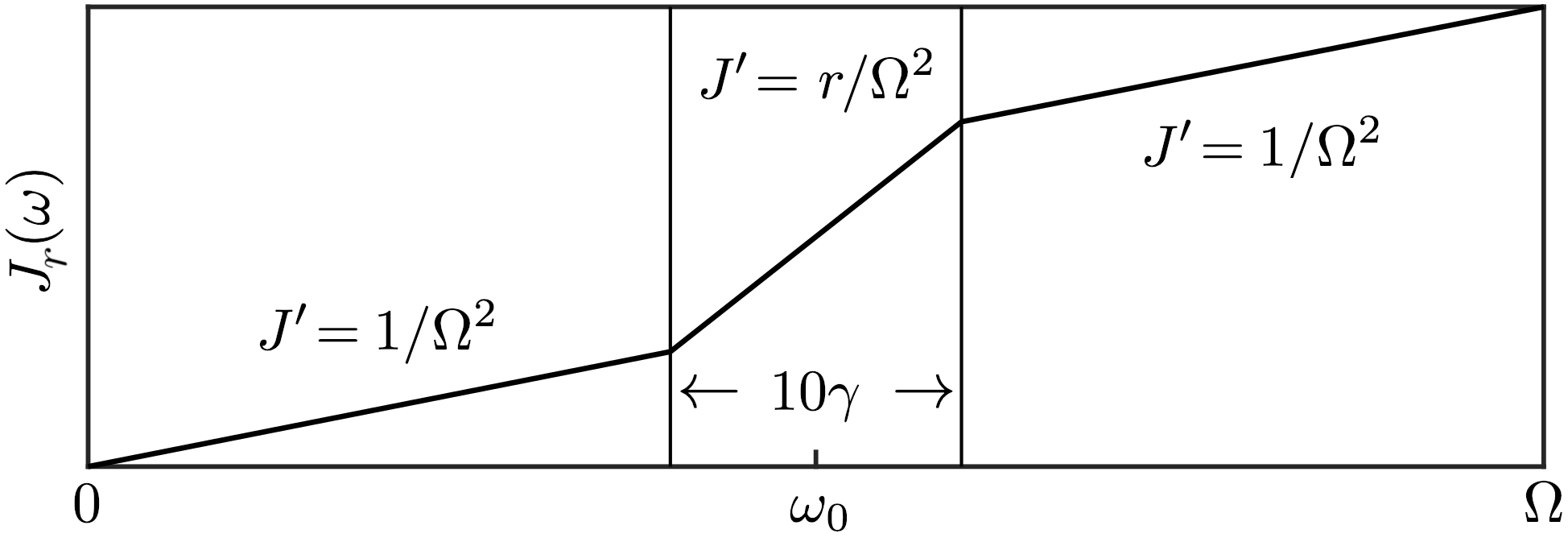}
\caption{(Color online) The piece-wise linear spectrum, $J_r(\omega)$, that we use to examine the effect of the slope of the spectral density on the accuracy of the master equations. On the two outer segments this spectrum  has the slope of the Ohmic spectrum, while on the middle segment the slope is increased by a factor $r$.
}
\label{fignewspec}
\end{figure}

If we use an Ohmic spectrum, then we cannot increase the slope of the spectral density without similarly increasing the decay rate(s). We already know that the master equations break down (deviates from the exact evolution) at large damping. To explore how the slope of the spectral density affects the master equations we thus thus require a new spectral density, and we use the one depicted in Fig.\ref{fignewspec}. This density, which we will denote by $J_r(\omega)$, is divided into three segments. It has a constant slope in each segment, with all but the middle segment having the same slope as the Ohmic spectrum ($dJ_r(\omega)/d\omega = 1/\Omega^2$). In the middle segment, which includes the transition frequency(ies) of the system, this slope is increased by a factor of $r$.  

It has already been established that the spectral density must be sufficiently flat in order that the damping induced by the bath be exponential~\cite{Santra17,Barnett03}. We can use system identification, already discussed in Section~\ref{SID}, to determine the number of dynamical variables required to reproduce the evolution of the open system as the slope of the spectral density is increased. When this number is greater than that possessed by the system, the evolution is non-Markovian and all time-independent Markovian master equations will break down. This allows us to determine not only how the B-R equation performs, but whether any Markovian master equation is able to model baths with steeply varying spectra. 

We examine the exact evolution of both a two-level system and the three-level V-system of Fig.\ref{6systems}a. The frequency of the two-level transition is $\omega_0 = 10\pi\tilde{\nu}$, with a damping rate of $\gamma = 0.296\tilde{\nu}$ when $r=1$, while the V-system has $\omega_1 = \omega_0 - \gamma/2$, $\omega_2 = \omega_0 + \gamma/2$, and $\gamma_1 = \gamma_2 = 0.2\tilde{\nu}$. Since we have chosen a higher damping rate for the two-level system the ``baseline" error of the master equations for this system will be a little higher than that for the V-system. 

\begin{figure}[t] 
\centering
\leavevmode\includegraphics[width=1\hsize]{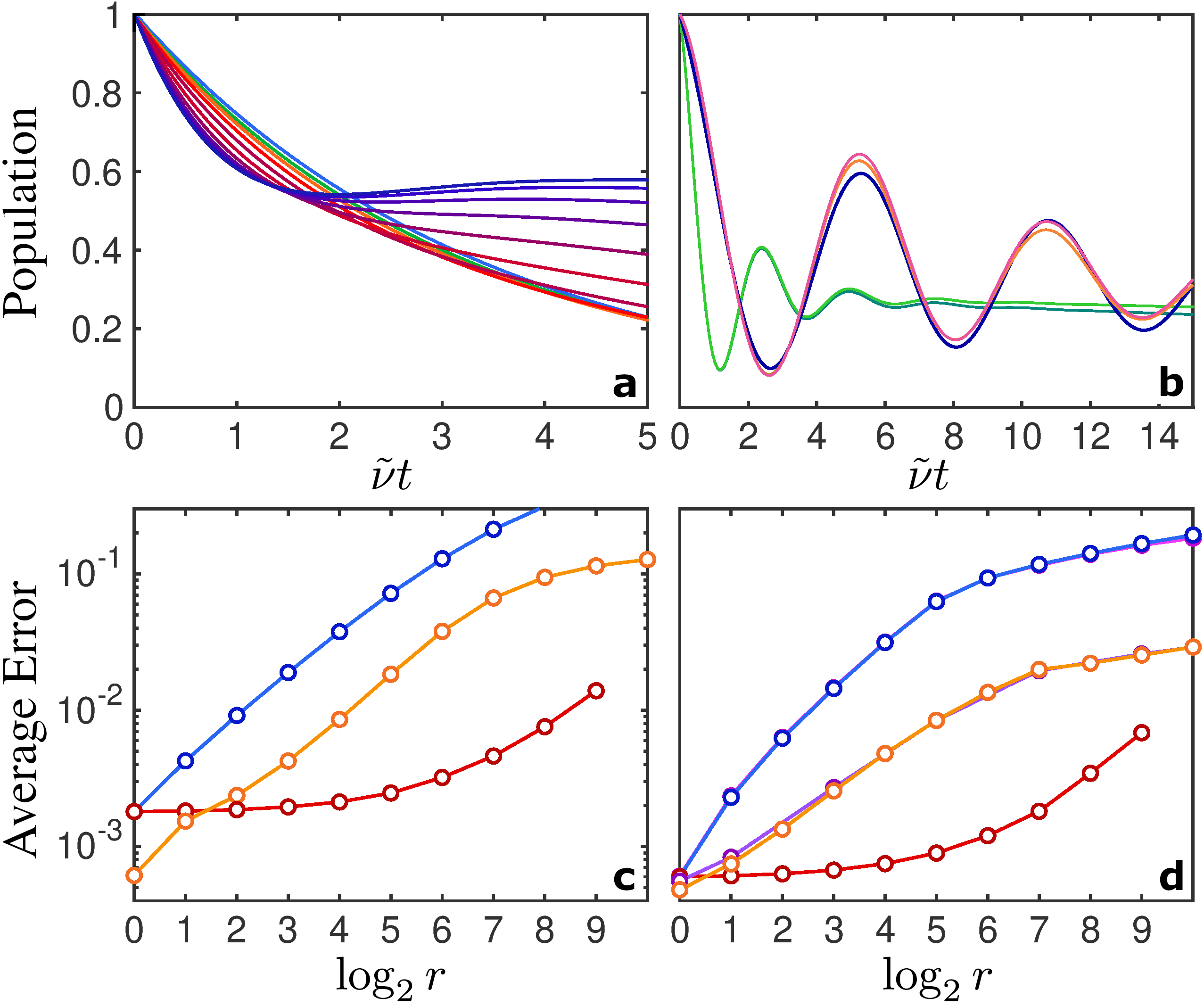}
\caption{(Color online) The breakdown of the Lindblad and Bloch-Redfield (B-R) master equations as the condition on the flatness of the spectral density is relaxed, for both a two-level system and the three-level V-system of Fig.\ref{6systems}a. a) The decay of the two level system from the excited state for increasing values of the spectral slope at the transition frequency, $J_r'(\omega_0) = r/\Omega^2$. The blue curve is the Ohmic spectrum, and rest of the curves, from green through violet, show the evolution for $r=2^n$, with $n=3, 4, \ldots, 12$. b) The decay of the V-system from state $|2\rangle$ at $r = 2^7$ (blue) with the predictions of the Lindblad and B-R master equations (green and turquoise, respectively). We also show the ``best-fit" to the exact dynamics for these master equations, obtained by fitting the parameters $R_j$ and $I_j$. c) The error of the single master equation for the two level system with spectrum $J_r$ as $r$ is increased (blue), and the error of the master equation for the Ohmic spectrum with the corresponding value of the damping rate, $\gamma(r)$ (red). We also show the error of the master equation for the spectrum $J_r$ obtained by fitting the damping rate. d) The respective errors of both master equations for damping from the initial state $|2\rangle$ for the spectrum $J_r$ (Lindblad: blue, B-R: magenta); the respective errors for the Ohmic spectrum with the equivalent damping rates (Lindblad: red, B-R: dark red); the respective errors for the ``best-fit" to the exact evolution, obtained by fitting the values of $R_j$ and $I_j$ separately for both master equations. Note that the difference between the errors of the two master equations are hardly discernible on the plot.
}
\label{figslopeerror}
\end{figure}

As we increase the slope of $J_r$, all the damping rates do increase (at first only a little) because the values of $J_r(\omega)$ at the transition frequencies also increase. The increase in the error of the master equations will thus have two sources, the increasing slope and the increasing damping rate(s). By determining the error of the master equations as we increase the damping rates while keeping the spectrum Ohmic, we can largely distinguish the relative contributions of the two sources of error. 

In Fig.\ref{figslopeerror}a we plot the population of the excited state of the two-level system as it decays into the bath, for $r = 2^n$, with $n = 0$ (the Ohmic spectrum) and $n = 3,4,\ldots, 12$. We see that as the slope increases the evolution distorts away from exponential decay. Since the system starts in the excited state, the coherences remain zero and there is only one independent variable in the evolution. Since a time-independent linear equation can only generate real non-exponential evolution if it has more than one dynamical variable, any deviation from exponential behavior necessarily implies non-Markovian evolution. In Fig.\ref{figslopeerror}c we plot the error in the evolution of the master equations (note that both the Lindblad and B-R master equations are identical for a two-level system) as a function of $n$ (equivalently $\log_2 r$). Note that the evolution of the master equation is determined solely by the damping rate $\gamma(r) = 2\pi g^2 J_r(\omega_0)$. Rather than using the damping rate as given by the master equation, for each value of $r$ we can alternatively choose the damping rate that minimizes the error. We also plot this ``best fit" error in Fig.\ref{figslopeerror}c, as well as the error of the master equations for an Ohmic spectrum with the damping rate $\gamma(r)$. We see that the effect of the increased slope on the error is much greater than that of the increased damping rate alone. We also see that the ``best-fit" error, while smaller than that given by the value of $\gamma$ specified by the master equation, increases just as rapidly as the latter. 

We use system identification to determine the dimension of the dynamics that generates the exact evolution for each value of $r$. We define this dimension as the number of dynamical variables required to account for $0.999$ of the combined magnitude of the dynamical eigenvalues (see Section~\ref{SID}). The results are presented in Table~\ref{dimTable}. We see that the dimension, and thus the non-Markovianity, starts increasing as soon as we increase the slope, showing that all time-independent master equations break down. 

\begin{table}[t]
\begin{tabular}{c|cccccccccccccc} 
$r$ & $2^0$ & $2^1$ & $2^2$ & $2^3$ & $2^4$ & $2^5$ & $2^6$ & $2^7$ & $2^8$ & $2^9$ & $2^{10}$ & $2^{11}$ & $2^{12}$ & $2^{13}$ \\
\hline
$D_2$ & 1 & 2 & 2 & 2 & 3 & 4 & 4 & 4 & 4 & 5 & 5 & 5 & 5 & 5 \\
\hline
$D_V$ & 4 & 5 & 6 & 7 & 8 & 9 & 10 & 10 & 11 & 11 & 12 & 12 & 12 & 12 \\
\end{tabular}
\caption{The effective dimension of the damped two-level system, $D_2$, and that of the damped three-level V-system, $D_V$, as the slope of the spectral density, $J'(\omega) = r/\Omega^2$, is increased for a fixed cut-off frequency, $\Omega$}
\label{dimTable}
\end{table}

We now examine whether there is difference between the accuracy of the Lindblad and B-R master equations as the slope, $r$, increases. For this we need to explore the evolution of the V-system. We choose the initial state of this system to be $|2\rangle$, and plot in  Fig.\ref{figslopeerror}d the error of both master equations as a function of $n$, as well as the error when $R_j$ and $I_j$ are chosen so as to give the best fit to the exact dynamics. As Fig.\ref{figslopeerror}d shows, the behaviour of the V-system is very similar to that of the two-level system. Even when both master equations have deviated significantly from the exact dynamics, they remain so close to each other that the difference in their respective errors is hardly distinguishable on the plot. Thus the master equations break down well before they deviate from each other, and thus well before the SVS approximation breaks down. We also note that the B-R equation continues to maintain positivity to good approximation up through $r=10$; the magnitude of the most negative eigenvalue of the density matrix remains below $1\times 10^{-12}$. 

In Fig.\ref{figslopeerror}b We show how the master equations compare to the exact evolution for the V-system when $n=7$ ($r=128$). Notable is how different the master equations are from the exact evolution, while being close to eachother. The fact that the master equations are able to approximate the exact evolution considerably better given the optimal choices for $R_j$ and $I_j$ is rather interesting. It suggests that there might be some way to develop improved formulae for $R_j$ and $I_j$. 

\begin{figure}[t] 
\centering
\leavevmode\includegraphics[width=1\hsize]{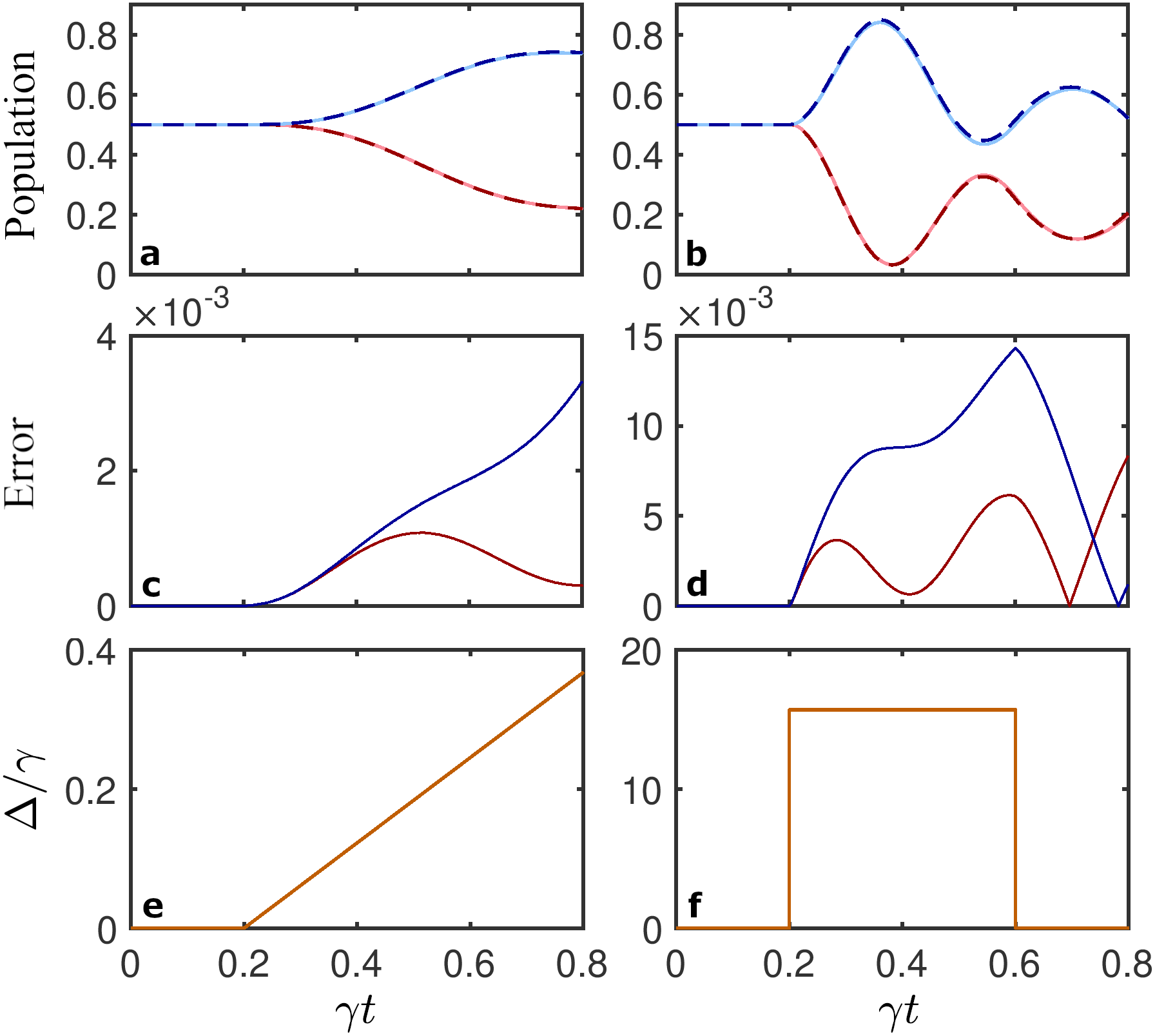}
\caption{(Color online) A comparison of the evolution predicted by the adiabatic extension of the new master equation, Eq.(\ref{HQM}), with an exact simulation of a V system coupled to an Ohmic bath (Fig.\ref{6systems}a), in which the detuning between the transitions, $\Delta\omega$, has the time-dependence given in Eqs.(\ref{F1}) and (\ref{F2}). The initial state is $|\psi_-\rangle = (|1\rangle - |2\rangle)/\sqrt{2}$ which is the dark state for degenerate transitions with equal damping rates. The transition frequency $\omega_1 = 3 \pi\tilde{\nu}$, the damping rates are $\gamma_1 = \gamma_2 = \gamma = 0.1\tilde{\nu}$, and the simulation time is $T=8$. (a,c,e) $\Delta\omega(t) = F_1(t)$ (Eq.(\ref{F1})). a) The populations of levels $|1\rangle$ (blue) and $|2\rangle$ (red), with the evolution of the adiabatic master equation denoted by dashed lines and that of the exact simulation by solid lines. (c) The absolute value of the difference between the populations predicted by the adiabatic master equation and the exact evolution. (e) The detuning as a function of time. (b,d,f) The same set of results but with the detuning given by $F_2(t)$  (Eq.(\ref{F2})).} 
\label{figDark}
\end{figure}

\section{Time-dependent problems: accuracy of the adiabatic extension} 
\label{t_apps}

Many important problems involve open systems whose Hamiltonians change with time. Our master equation can be used to describe these systems if the time-dependence is not too fast. To do so one takes the master equation and changes the parameters and operators that appear in it, namely $\gamma_j(\omega_j)$, $\Delta_j(\omega_j)$, and $\sigma_j$ (which depend on the system eigenstates and thus on the system Hamiltonian), so that at each time they take the values determined by the Hamiltonian of the system at that time~\cite{McCauley20}. The resulting time-dependent ``adiabatic'' master equation will be effective for sufficiently slow changes in the Hamiltonian. 

Here we examine some examples to confirm that the adiabatic version of the master equation is accurate even when two levels of an open system cross each other, or move from degenerate to near-degenerate, during the evolution. We consider first the V system in which both damping rates are equal and the detuning changes with time. We start the system in the state $|\psi_-\rangle \equiv (|1\rangle - |2\rangle)/\sqrt{2}$, which for $\Delta\omega = 0$ will not decay since it is the (sub-radiant) dark state~\cite{Gross82,Stannigel12}. We then change the detuning with time as determined by following two functions: 
\begin{align}
      F_1(t) & =  \left\{ \begin{array}{rl}
         0 , & \;\; 0 <  t < \frac{T}{4} ,\\[1ex]
         \frac{\pi}{64} \left(t-\frac{T}{4}\right) ,    & \;\; \frac{T}{4} <  t < T ,
      \end{array}  \right. \label{F1} \\[1ex]
     F_2(t) & =  \left\{ \begin{array}{rl}
         0 , & \;\; 0 <  t < \frac{T}{4} ,\\[1ex]
        \frac{\pi}{2},    & \;\; \frac{T}{4} <  t < \frac{3T}{4}, \\[1ex]
         0 , &  \;\;  \frac{3T}{4}  < t < T . 
      \end{array}  \right.   \label{F2}  
\end{align}
The function $F_1$ is chosen so that the detuning increases gradually, while $F_2$ involves rapid changes. In Fig.\ref{figDark} we compare the adiabatic version of the master equation with the exact evolution for the two cases. For $\Delta\omega(t) = F_1(t)$ the maximum error of the adiabatic master equation is $3.4 \times 10^{-3}$, and for $\Delta\omega(t) = F_2(t)$ the maximum error is  $1.5 \times 10^{-2}$. 

\begin{figure}[t] 
\centering
\leavevmode\includegraphics[width=0.93\hsize]{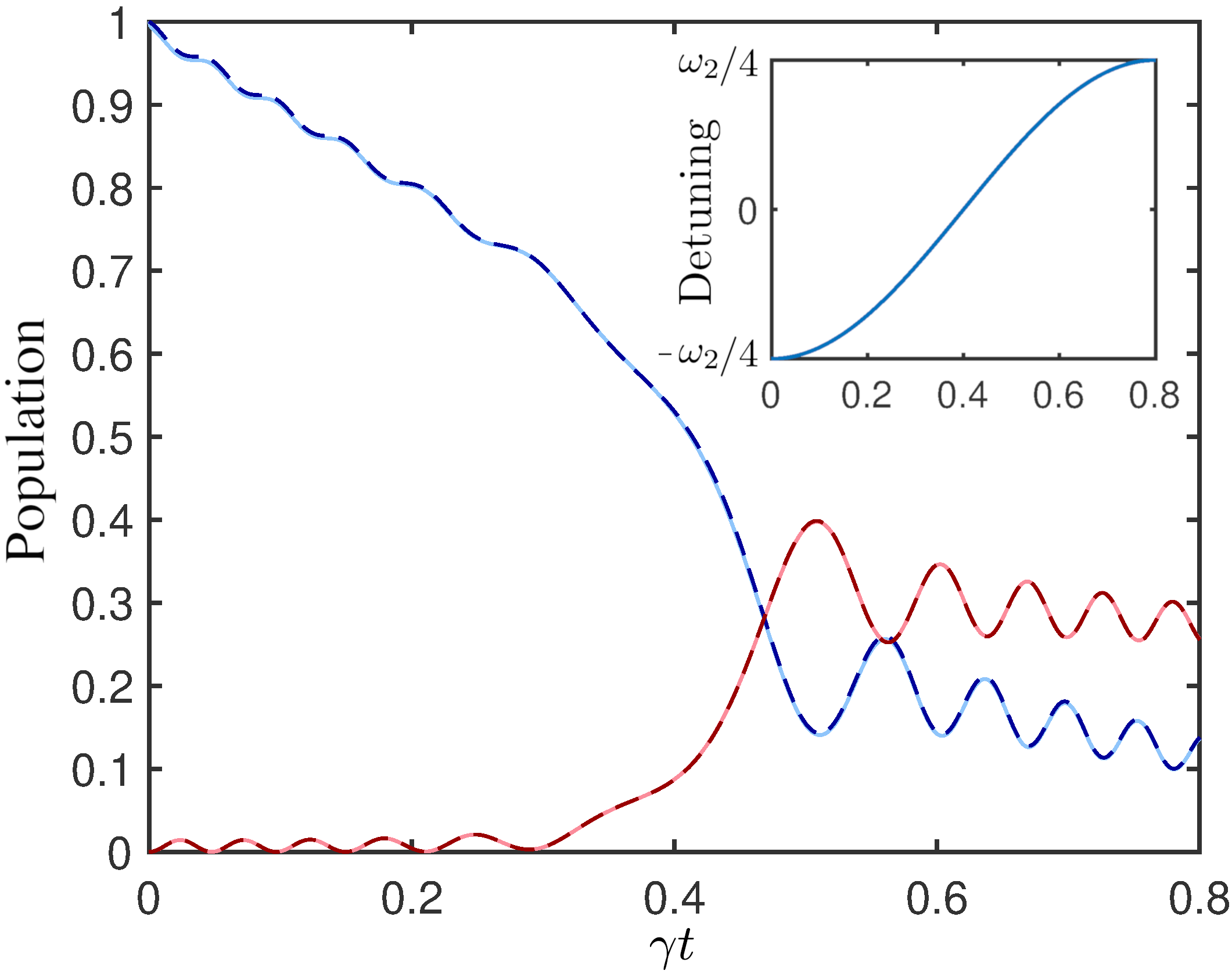}
\caption{(Color online) Here we plot the evolution resulting from a generalized Landau-Zener transition in which the energies of two coupled levels cross. The levels are the upper two levels of the four-level system depicted in Fig.\ref{6systems}c. The energy of level $|2\rangle$ is fixed so that $\omega_2 = 2\pi\tilde{\nu}$, and that of level $|1\rangle$, in which all the population starts, increases with time. The detuning between the levels is shown as a function of time in the inset. We plot the populations of the two levels, both the exact evolution (solid) and that predicted by the master equation (dashed). The damping rates of the two levels are $\gamma_2 = 2\gamma_1 = 0.05\tilde{\nu}$, the coupling between them is $c = 0.2\tilde{\nu}$, the initial detuning is $\Delta_0 = (\pi/2)\tilde{\nu}$, and the period of the sinusoid is $2\pi/\nu = 64/\tilde{\nu}$. 
}
\label{figLZ}
\end{figure}

As our final example we consider a generalized version of the Landau-Zener transition~\cite{Wittig05,Wubs06}, in which the energies of two coupled levels cross each other. In particular we use the 4-level system depicted in Fig.\ref{6systems}c, in which we add a coupling between the upper two levels. In the original Landua-Zener transition, for which there is an analytic solution, the energy of one of the levels is fixed and the other increases linearly with time. We generalize this by choosing the following sinusoidal time-dependence for $\omega_1$: 
\begin{align}
      \omega_1(t) = \omega_2 - \Delta_0 \cos(\nu t)  , \;\;\;\; 0 \leq t \leq \smallfrac{\pi}{\nu} .  \label{w1t}
\end{align}
The detuning, $\Delta\omega \equiv \omega_1 - \omega_2$, starts at $-\Delta_0$ and increases as a sinusoid through zero to end at  $\Delta_0$. We plot the evolution of the populations of the two levels in Fig.\ref{figLZ}. The maximum error of the adiabatic extension of the master equation is less than $5.4 \times 10^{-3}$. 

\section*{Discussion}
\label{conc}

We have shown that a slow variation condition on the spectral density is necessary for Markovianity, and along with weak damping and a high cut-off frequency is a sufficient condition for the existence of an accurate Lindblad master equation for all regimes of detuning. In doing so we have shown that the Bloch-Redfield equation can be replaced with this Lindblad equation. This resolves the long-running controversy with the Bloch-Redfield equation, and confirms the conjectures of Eastham \textit{et al.}~\cite{Eastham16}.


The new master equation unifies the existing Linblad master equations for degenerate and non-degenerate systems and in doing so provides insight into the dynamics of the near-degenerate regime. It also allows both the use of efficient Monte Carlo methods and a measurement description of the action of a thermal bath for all regimes. Further, its adiabatic extension provides a powerful tool for simulating systems in which transitions are time-dependent and cross during the evolution, so long as this time-dependence is not too fast. This suggests that further exploration of the accuracy of the adiabatic extension as a function of the speed of the time-dependence may be a worthwhile endeavor. Such an exploration would help to delineate the class of controlled systems for which it is effective. 

The technique of system identification played an important role in obtaining the master equation, as well as determining when open systems are Markovian. As far as we are aware, system identification has not been used before as a tool to understand the dynamics of open quantum systems, or emergent phenomena in many-body systems more generally. We expect that it will prove to be powerful for exploring a wide range of problems in open systems and many-body physics. 

\section*{Acknowledgements} The authors thank Even Jensen for graphic design of Fig.\ref{6systems}. This research was supported in part by appointments to the Student and Postgraduate Research Participation Programs at the U.S Army Research Laboratory~\footnote{These programs are administered by the Oak Ridge Institute for Science and Education and the Oak Ridge Association of Universities through an interagency agreement between the U.S Department of Energy and USARL.}. D.I.B. was supported by Air Force Office of Scientific Research Young Investigator Research Program (FA9550-16-1-0254). 


\appendix 

\section{System identification for linear systems} 
\label{App:SID} 

Here we present the method we used to determine the minimal model of a linear system given a knowledge of the evolution of a subset of the system's state-space. In our case the linear system consists of a low-dimensional quantum system interacting with a high-dimensional bath, and the subset of the state space that we can observe is the density matrix of the low-dimensional system. The following method is one of a family of elegant methods referred to as \textit{subspace identification methods}, adapted so as to use a set of initial conditions rather than a set of inputs. Further information on subspace identification methods can be found  in~\cite{Katayama05,Overschee96,Keesman11,Ruscio95}. 

Let us say we have a high dimensional system with dimension $J$ (in our case the open system and the bath), and we have the ability to observe $N < J$ variables of the system, as well as to evolve the system with any choice of initial conditions for the $N$ variables we can observe. We would like to find an accurate model (another linear system) that generates the evolution of the $N$ variables but is only $M$ dimensional with $N \leq M  < J$. Let us denote a state of the full $J$ dimensional system by the vector $\mathbf{v}$, and the subset of $N$ variables in which we are interested by the $N$-dimensional vector $\mathbf{x}$. The map that gives the state of the total $J$-dimensional system at time $\tau$ given an initial state $\mathbf{v}(0)$ we will call $Z(\tau)$ so that $\mathbf{v}(\tau) = Z(\tau)\mathbf{v}(0)$. Defining $Z \equiv Z(\tau)$ we note that $Z(n\tau) = Z^n$. We also define the non-square projector $P$ that projects onto the $N$ variables so that $\mathbf{x} = P \mathbf{v}$. 

Given the ability to evolve the total system with any choice of initial conditions for the $N$-dimensional subsystem, along with a single choice for the initial values of the rest of the variables (of which there are $N-J$), we can obtain the matrices $Y_n = P Z(n\tau) P^{\ms{T}}$ that maps the $N$ variables at time $0$ to their values at time $\tau$, for any time $\tau$.

We now construct the following two symmetric ``block Hankel'' matrices: 
\begin{align}
 H_0 & \equiv \left(
\begin{array}{cccc}
Y_0  & Y_1  &  \cdots  & Y_n  \\
Y_1  &  Y_2 &    & Y_{n+1}  \\
 \vdots  &   & \ddots  &   \vdots \\ 
 Y_n     &  Y_{n+1} & \cdots  &  Y_{2n}
\end{array}
\right),  \\
 H_1 & \equiv  
 \left(
\begin{array}{cccc}
Y_1  & Y_2  &  \cdots  & Y_{n+1}  \\
Y_2  &  Y_3 &    & Y_{n+2}  \\
 \vdots  &   & \ddots  &   \vdots \\ 
 Y_{n+1}   &  Y_{n+2} & \cdots  &  Y_{2n+1}
\end{array}
\right).
\end{align} 
We now note that $H_0$ can be written as an outer product $H_0 = C D^{\ms{T}}$ of (non-square) matrices given by 
\begin{align} 
 C & = \left( \begin{array}{c} P   \\ P Z  \\ \vdots  \\  P Z^{m} \end{array} \right) = P \left( \begin{array}{c} I   \\  Z  \\ \vdots  \\   Z^{m} \end{array} \right) , \\
  D^{\ms{T}}  & = \left(   \;  Z \; Z^2 \; \cdots  \;  Z^{m} \right) P^{\ms{T}}. 
\end{align}
We can now determine $C$ and $D$ by doing a singular value decomposition of $H_0$ to give $H_0 = U S V^{\ms{T}}$. Note that since the smaller dimension of the matrices $C$ and $D$ is smaller than that of $H_0$ we expect many of the columns of U and the rows of $V^{\ms{T}}$ will be zero, as will many of the eigenvalues of $H_0$ which are given in the diagonal matrix $S$. Note that we can now view $P$ and $Z$ as defining a linear model that generates evolution of the $N$-dimensional subsystem. The number of eigenvalues that are appreciably non-zero tells us the dimension of the model. 

To distinguish the model from the original total system we started with, we can write the matrices $C$ and $D$ as 
\begin{align} 
 C^{\ms{T}} & = \left(   \;  M^{\ms{T}} \; [M^{\ms{T}}]^2 \; \cdots  \;  [M^{\ms{T}}]^m \right) Q^{\ms{T}}, \\
  D^{\ms{T}}  & = \left(   \;  M \; M^2 \; \cdots  \;  M^{m} \right) Q^{\ms{T}} , 
\end{align}
where $M$ is the evolutionary map for the model and $Q$ is the projector onto the subsystem. Let us now decompose $H_0 = U S V^{\ms{T}}$ into the outer product of two vectors $\tilde{C} \equiv U \sqrt{S} P^{\ms{T}}$ and $\tilde{D} = (P \sqrt{S} V^{\ms{T}})^{\ms{T}}$. Noting that  
\begin{equation}
     H_1 = C Z D^{\ms{T}}   
\end{equation}
we can obtain the evolutionary map for the model, $M$, from $H_1$ using  
\begin{equation}
     M =  (\tilde{C}\tilde{C}^{\ms{T}})^{-1} \tilde{C}^{\ms{T}} H_1 \tilde{D} (\tilde{D}^{\ms{T}}\tilde{D})^{-1} . 
\end{equation}
We note that the above method determines the equations of motion of the linear system, but does not itself give the change of basis between the original system variables and those that appear in the obtained equations. Obtaining this change of basis requires additional methods. 

\section{Principle value of an integral} 
\label{PV}

In deriving the master equation in Section~\ref{derivME} we used the fact that 
\begin{align}
    \int F(x) \left[ \int_0^\infty \!\!\! \sin([\omega - \omega_0]s) ds\right] dx = \mathbb{P} \left[ \int \frac{F(x)}{x} dx  \right] 
\end{align}
for any smooth function $F(x)$, in which $\mathbb{P}\left[ \cdots \right]$ denotes the \textit{principle value} of a divergent integral. 
The principle value of an integral that diverges at a point $a$ (where $a \in (b,c)$), is defined by 
\begin{align} 
    \mathbb{P}\left[ \int_b^c f(x) dx  \right] & \equiv \lim_{\varepsilon\rightarrow 0}  \left[ \int_b^{a-\varepsilon} \!\!\!\!\!\! \!\! f(x) dx + \int_{a+\varepsilon}^c \!\!\!\!\!\! f(x) dx \right] . 
\end{align}
Since the divergent function $f(x) = 1/x$ is anti-symmetric and diverges at $a = 0$, it is simple to evaluate the principle value of $\int_{-b}^c (1/x) dx$. Assuming that $b$ and $c$ are positive and $c > b$ we have 
\begin{align} 
    \mathbb{P}\left[ \int_{-b}^c \frac{dx}{x}   \right] & \equiv \lim_{\varepsilon\rightarrow 0}  \left[ \int_{-b}^{-\varepsilon}  \frac{dx}{x} + \int_{\varepsilon}^c \frac{dx}{x} \right] \nn \\ 
    & = \lim_{\varepsilon\rightarrow 0}  \left[ \int_{-b}^{-\varepsilon} \frac{dx}{x}+ \int_{\varepsilon}^b \frac{dx}{x} \right]  + \int_b^c \frac{dx}{x} \nn \\
    & = \int_b^c \frac{dx}{x} = \ln(c/b). 
\end{align}

\vspace{3ex}
\noindent \textbf{Competing interests: } 
The authors declare no competing financial interests. 

\noindent \textbf{Data availability statement:} 
All numerical results (datasets) generated and analyzed in this work can be obtained from the corresponding author upon reasonable request. 

\noindent \textbf{Code availability statement:} 
All computer code used to generate and/or analyze the numerical results presented in this work can be obtained from the corresponding author upon reasonable request. 


%

\end{document}